\documentclass[pdflatex,sn-mathphys-num]{sn-jnl}% Math and Physical Sciences Numbered Reference Style
\usepackage{graphicx}%
\usepackage{multirow}%
\usepackage{amsmath,amssymb,amsfonts}%
\usepackage{amsthm}%
\usepackage{mathrsfs}%
\usepackage[title]{appendix}%
\usepackage{xcolor}%
\usepackage{textcomp}%
\usepackage{manyfoot}%
\usepackage{booktabs}%
\usepackage{algorithm}%
\usepackage{algorithmicx}%
\usepackage{algpseudocode}%
\usepackage{listings}%
\usepackage[percent]{overpic}
\theoremstyle{thmstyleone}%
\theoremstyle{thmstyletwo}%

\theoremstyle{thmstylethree}%

\begin{document}

\title[]{Reducing Ion Heating in Quantum Computing: A Novel 3D-Printed Micro Ion Trap with Skeleton Structure}


\author*[1]{\fnm{Chon-Teng Belmiro} \sur{Chu}}\email{belmirochu@gmail.com}

\author[1]{\fnm{Hao-Chung} \sur{Chen}}%\email{iiauthor@gmail.com}

\author[1,2,3]{\fnm{Ting} \sur{Hsu}}%\email{iiiauthor@gmail.com}

\author[1]{\fnm{Hsiang-Yu} \sur{Lo}}

\author[1,4,5]{\fnm{Ming-Shien} \sur{Chang}}

\author*[1,2,3]{\fnm{Guin-Dar} \sur{Lin}}\email{guindar.lin@gmail.com}

\affil[1]{\orgdiv{Trapped-Ion Quantum Computing Laboratory}, \orgname{Hon Hai Research Institute}, \orgaddress{ \city{Taipei}, \postcode{11492}, \country{Taiwan}}}

\affil[2]{\orgdiv{Department of Physics and Center for Quantum Science and Engineering}, \orgname{National Taiwan University}, \orgaddress{ \city{Taipei}, \postcode{10617},  \country{Taiwan}}}

\affil[3]{\orgdiv{Physics Division}, \orgname{National Center for Theoretical Sciences}, \orgaddress{ \city{Taipei}, \postcode{10617},  \country{Taiwan}}}

\affil[4]{\orgdiv{Institute of Atomic and Molecular Sciences}, \orgname{Academia Sinica}, \orgaddress{ \city{Taipei}, \postcode{10617},  \country{Taiwan}}}

\affil[5]{\orgdiv{Department of Physics and Center for Quantum Technology}, \orgname{National Tsing-Hua University}, \orgaddress{ \city{Hsinchu}, \postcode{30013},  \country{Taiwan}}}


\abstract{
Electric-field-induced ion heating is a major obstacle in scalable trapped-ion quantum computing. 
We present a theoretical study of a novel 3D-printed ion trap with a skeleton electrode structure, designed to reduce heating by minimizing surface area near the ion.
Compared to a conventional blade trap with identical confinement parameters, the skeleton trap achieves over 50\% reduction in total heating rate.

Patch-by-patch analysis reveals that heating is dominated by surfaces within 500~$\mu$m of the ion. 
For axial motion, the peak heating occurs approximately 110~$\mu$m away due to electric field directionality. 
%We demonstrate that minor geometric optimization—realigning electrode gaps with these hotspots—can further suppress heating, despite increased surface area. 
We demonstrate that minor geometric optimization, in which the electrode gaps are realigned with these hotspots, can further suppress heating despite the associated increase in surface area.
A linear relationship between ion-to-electrode distance and peak heating location is also established.

These results highlight the potential of 3D-printed electrode designs for achieving both strong confinement and reduced noise in future quantum systems.
}

\maketitle

\section{Introduction}\label{sec1}

Trapped-ion systems have emerged as a leading platform for quantum computing, offering long coherence times, high-fidelity gate operations, and all-to-all qubit connectivity \cite{Wineland1998, Haffner2008, Blatt2008, Brown2011, Cho2015,  Bruzewicz2019, Png2022}.
In these systems, ions are confined by electromagnetic potentials in radio-frequency (RF) Paul traps, where qubit states are encoded in long-lived internal energy levels. The ability to isolate ions from environmental noise while maintaining precise control makes ion traps ideal candidates for fault-tolerant quantum processors \cite{Monroe2014, Postler2022}.

Traditionally, macroscopic 3D Paul traps with millimeter-scale electrodes have been widely used due to their deep and harmonic pseudopotentials. These geometries are well-suited for quantum logic operations, fast and efficient Doppler cooling, and laser-based state manipulation \cite{Leibfried2003}. However, as the development of multi-qubit devices accelerates and researchers aim for large-scale quantum architectures, these traps face practical challenges related to fabrication limitations and scalability. 
To overcome these limitations, 
researchers have turned to planar surface-electrode traps that enable microfabrication and integration into multi-qubit arrays \cite{Kielpinski2002, Chiaverini2005, Seidelin2006, Pino2021, Moses2023, Chen_2024}. 
While such devices are scalable, bringing ions closer to the electrode surface exacerbates motional heating caused by fluctuating electric fields, particularly those arising from microscopic patch potentials \cite{Turchette2000,Hite2012,Safavi-Naini2011,Daniilidis2014}.
The heating rate typically scales with the ion-to-electrode distance as $d^{-\alpha}$ with $\alpha \approx 4$, placing a fundamental constraint on trap miniaturization \cite{Deslauriers2006,Dubessy2009}. 
As a result, despite their integration benefits, surface traps suffer from elevated heating rates that limit gate fidelity and reduce coherence.

To address these trade-offs, a number of groups have developed monolithic and integrated 3D trap architectures that circumvent conventional assembly challenges.
In particular, the work of Auchter et al. demonstrated an ion trap fabricated via wafer stacking and photolithography, yielding a monolithic device with a trap depth exceeding 1~eV, along with microfabricated alignment features for large-scale integration \cite{Auchter2022}.
Takahashi et al. at OIST fabricated 3D ion traps in fused silica using femtosecond laser etching, with gold-coated surfaces forming the electrodes. Their design incorporated etched cutouts to hold optical fibers with micro-mirrors at the tips, forming a stable Fabry-P\'{e}rot cavity around the ion.
This monolithic integration enabled strong ion-photon coupling for quantum networking experiments \cite{teh2024}.
Similarly, the group led by Pagano at Rice University demonstrated a monolithic linear ion trap with integrated wiring layers and precise through-chip via connections, illustrating a scalable approach to dense ion trap arrays \cite{Zhuravel2024MonolithicTrap}. These works, while not 3D printed, highlight the growing interest in advanced microfabrication techniques that push beyond the constraints of conventional machining.

More recently, 3D-printed micro ion traps have emerged as a particularly promising direction.
Quinn et al. introduced the "trench trap" using two-photon polymerization, achieving a hybrid planar-3D structure that maintained compactness while significantly enhancing pseudopotential harmonicity \cite{Quinn2022}.
Building on this approach, Xu et al. at UC Berkeley and Lawrence Livermore National Laboratory developed a fully 3D-printed microtrap that successfully confined a single $^{40}$Ca$^+$ ion at just 130~$\mu$m above the substrate with radial frequencies exceeding 20~MHz.
Their device delivered strong confinement and high-fidelity qubit operations using Doppler cooling alone, demonstrating that 3D-printed geometries can match or exceed the performance of planar traps \cite{Xu:2025aa}.

Motivated by these advances, our work introduces a new 3D-printed ion trap design featuring a hollow "skeleton" geometry that suppresses heating by minimizing electrode surface area near the ion while preserving strong harmonic confinement. Our design takes advantage of recent progress in metal additive manufacturing, made possible by collaboration with industrial partners such as \textit{Additive Intelligence}, who offer AI-enhanced 3D printing capabilities for high-precision electrode structures \cite{AdditiveIntelligence}.
This approach aims to suppress electric-field noise by spatially separating the ion from fluctuating patch potentials, without compromising trap depth or stability.

To assess its performance, 
we present a comparative theoretical analysis of the heating behavior in the skeleton trap versus a conventional blade trap,
using identical confinement parameters and ion-to-electrode distances. 
Our mode-resolved and spatially resolved patch analysis reveals that the skeleton geometry reduces heating by more than 50\%,
with further improvements possible through geometric optimization of electrode placement. 
%Notably, we show that axial-mode heating—governed by the directionality of electric field vectors—can be mitigated by aligning inter-electrode gaps with noise-dominant regions, 
%even if the overall surface area increases.
Notably, we show that axial-mode heating, which arises due to the directional characteristics of the electric field vectors, can be mitigated by aligning inter-electrode gaps with noise-dominant regions. This mitigation remains effective even in scenarios where the total electrode surface area is increased.

This study not only validates the potential of 3D-printed ion trap technology but also provides a clear pathway toward optimized trap geometries for next-generation quantum devices.
The structure of this paper is as follows: 
Section 2 reviews the physical principles of ion heating in Paul traps. 
Section 3 introduces the design methodology of our trap geometries. 
Section 4 presents the theoretical framework for calculating electric field noise and heating rates. 
%Section 5 analyzes and compares the results between trap configurations. 
Section 5 presents the heating rate simulations and design optimizations.
Finally, Section 6 concludes with a discussion of future implications and design considerations.

\section{Principles of Ion Heating in RF Paul Traps}\label{sec2}

Trapped ions are confined using electromagnetic potentials that produce a three-dimensional harmonic well, allowing precise control of their motional and internal quantum states.
In radio-frequency (RF) Paul traps, this confinement is achieved by a combination of static and RF electric fields.
The ions experience a time-averaged pseudopotential that results in stable harmonic confinement near the trap center \cite{Paul1990}.
However, 
%fluctuations in the surrounding electric fields
electric field noise,
particularly that originating from 
patch potentials on electrode surfaces,
can couple to the ion’s motion and induce heating and decoherence \cite{Wineland1998,Turchette2000,Safavi-Naini2011}.

\subsection{Harmonic Confinement and Noise Coupling}

For small oscillations around the trap minimum, the motion of a single ion can be modeled as a quantum harmonic oscillator along each principal axis.
The Hamiltonian for the ion motion is
\begin{equation}
    H_0=\hbar \omega_k \left( \hat{a}^\dagger_k \hat{a}_k+ \frac{1}{2} \right )
\end{equation}
where $\omega_k$ is the secular frequency in direction 
$k \in \{ x, y, z \}$, and $\hat{a}_k^\dagger$, $\hat{a}_k$ are the corresponding  creation and annihilation operators.
This description underpins the quantization of ion motion and the basis for laser cooling and quantum logic gates \cite{Leibfried2003}.

In the presence of a fluctuating electric field 
$\mathbf{E} (t)$,
the ion experiences a perturbing potential
$H'(t) = e \mathbf{E}(t) \cdot \mathbf{r}$,
which can induce transitions between motional states.
If the ion is initially cooled to the ground state, heating is defined as the rate at which energy is transferred from the field noise to the ion's motion.
This is typically measured as the rate of increase in the mean phonon number $\langle n_k(t) \rangle $ \cite{Turchette2000}.

\subsection{Quantum Treatment of Heating}

Using first-order time-dependent perturbation theory, 
the transition rate from the ground state $|0\rangle$ to the first excited state $|1\rangle$ in mode $k$ due to electric field noise is given by:
\begin{equation} \label{eq: transition rate from 0 to 1}
    \Gamma_{k}=
\frac{e^2}{4 m \hbar \omega_k}
S_{E_k} (\omega_k)
\end{equation}
where $m$ is the ion mass and $S_{E_k} (\omega_k)$ is the power spectral density of the electric field noise component $E_k(t)$ evaluated at the trap frequency $\omega_k$ \cite{Wineland1998}.
This expression establishes that ion heating is directly proportional to the local electric field noise at the ion’s position.
The spectral density is defined as
\begin{equation} \label{eq: the power spectral density of the fluctuating electric field}
    S_{E_k} (\omega_k) \equiv
2 \int_{-\infty}^{\infty}
\, d\tau 
e^{i \omega_k \tau}
\langle
E_k (t) E_k(t+\tau) 
\rangle
\end{equation}
which characterizes the temporal correlations of the field fluctuations along the mode direction $k$ \cite{Turchette2000,Brownnutt2015}.

\subsection{Surface Noise and Patch Potentials}
One of the dominant sources of electric field noise in ion traps arises from microscopic fluctuations on electrode surfaces, commonly known as patch potentials. 
These are believed to originate from inhomogeneous surface properties such as adsorbates, grain boundaries, or surface oxides, leading to time-varying microscopic dipole fields \cite{Labaziewicz2008,Safavi-Naini2011,Brownnutt2015,Hite_2021,Hite2017}.
The effect of these patch potentials becomes more pronounced as the ion is brought closer to the electrode surface, typically causing the heating rate to scale as $d^{-\alpha}$,
where $d$ is the ion-to-electrode distance and 
$\alpha$ ranges between 3 and 4 in most experimental observations \cite{Deslauriers2006}.

Understanding and modeling the spatial and spectral structure of this noise is therefore essential for designing ion traps with reduced heating.
In the next section, we will present the theoretical framework used to compute the electric field fluctuations and their contribution to heating, using Green’s functions and patch-resolved surface integrals \cite{Low2011}.

\section{Design of the 3D-Printed Skeleton Trap}\label{sec3}

In this work, we present a theoretical design and performance analysis of a novel 3D-printed ion trap featuring a skeleton electrode structure.
This design is intended to suppress ion heating by reducing the area of electrode surface directly exposed to the ion.
Our study provides a comprehensive simulation-based comparison with a conventional blade trap of identical geometric parameters to isolate the effects of electrode geometry on heating behavior.

\subsection{Geometric Configuration}

%Figure \ref{fig: traps} illustrates the schematic diagrams and side views of both the blade trap and the 3D-printed skeleton trap.
%Both traps are designed with an identical ion-to-electrode distance of 200 $\mu$m, defined as the shortest distance between the ion and the nearest point on any electrode surface.
%This choice is motivated by fabrication constraints in blade traps, where further reduction would necessitate impractically thin endcap electrodes, potentially compromising mechanical stability.
%Using the same distance in the 3D-printed trap ensures that any differences in performance, particularly heating rate, are attributable to trap geometry rather than differences in confinement strength.

%The blade trap uses a conventional quadrupole arrangement of four opposing blade-shaped electrodes.
%In contrast, the 3D-printed trap adopts a skeleton geometry that strategically removes bulk electrode material while preserving radial symmetry.

\begin{figure}[htbp]
    \centering
    % Top row
     \begin{overpic}[width=0.45\textwidth]{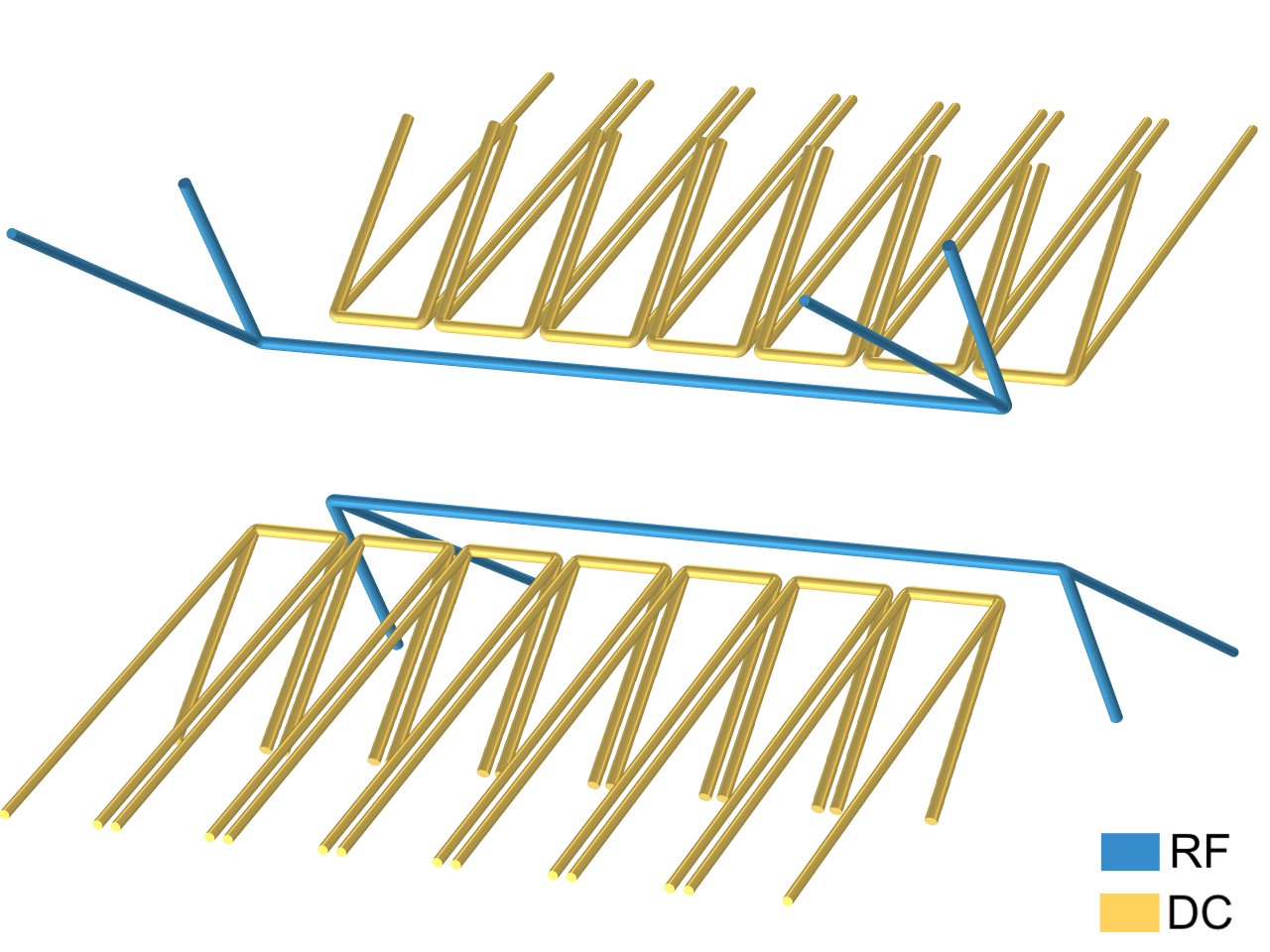}
        \put(0,76){a)}
    \end{overpic}
    \hfill
    \begin{overpic}[width=0.45\textwidth]{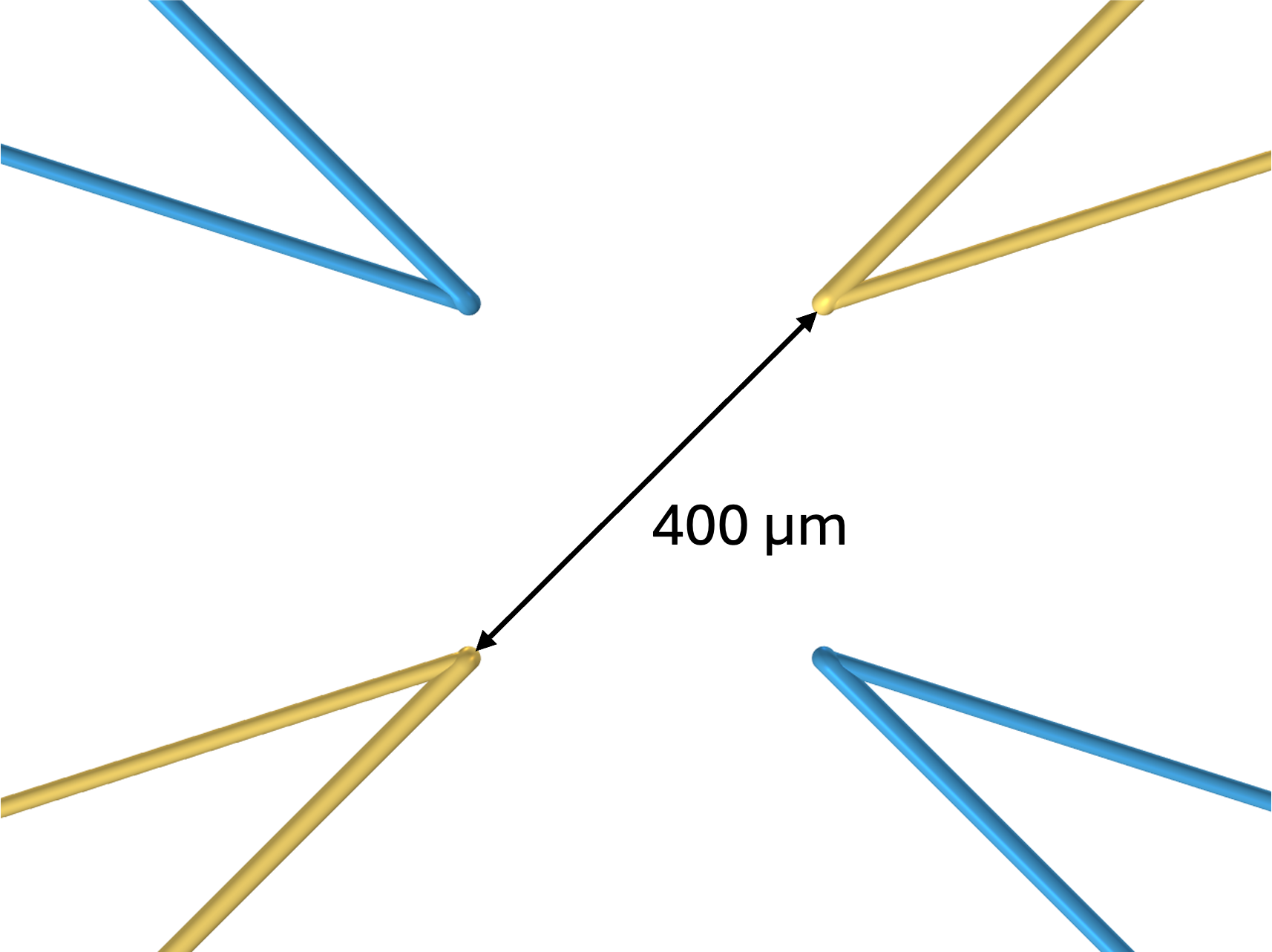}
        \put(0,76){ b)}
    \end{overpic}

    \vspace{0.5cm} % space between rows

    % Bottom row
    \begin{overpic}[width=0.45\textwidth]{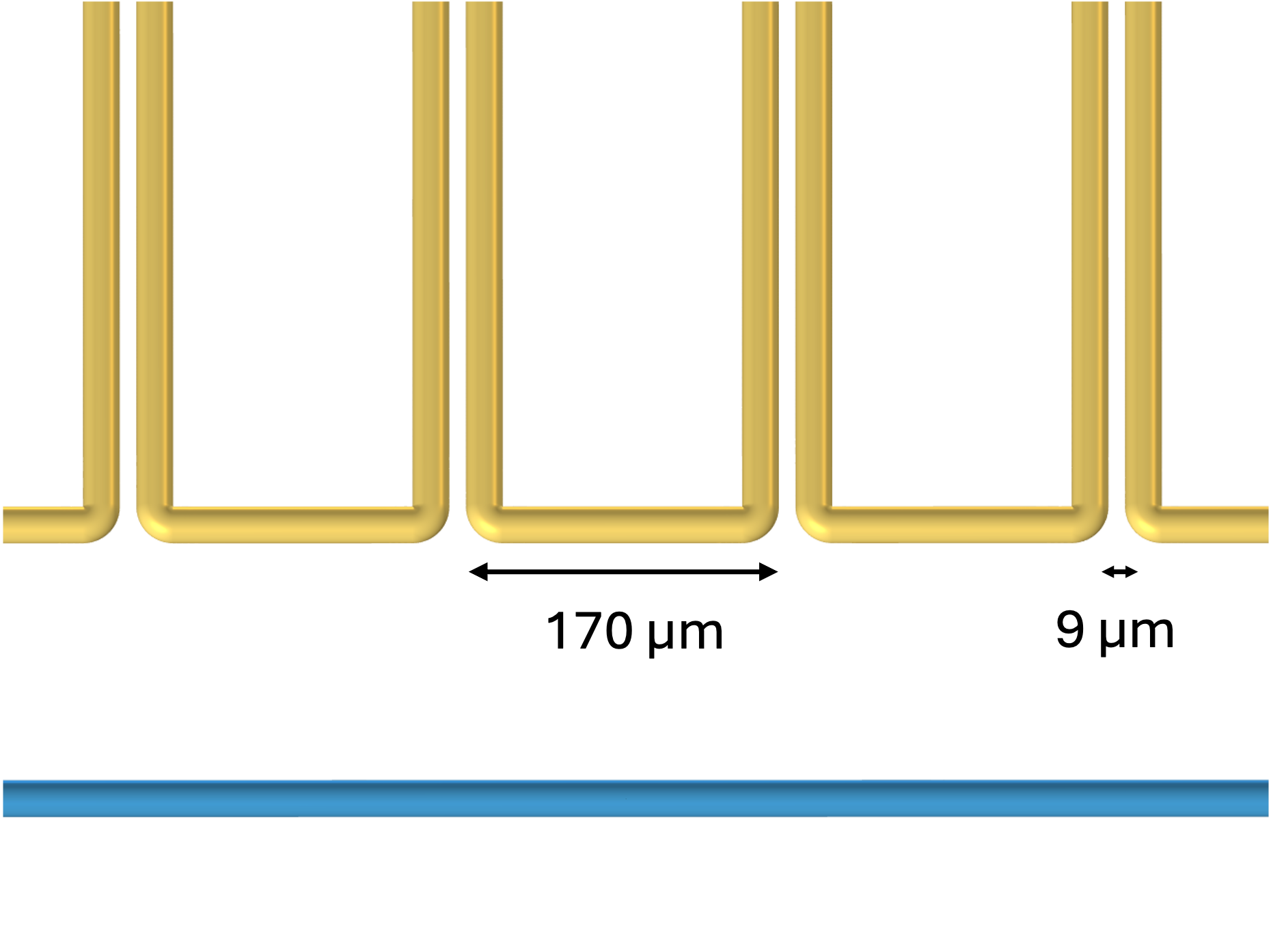}
        \put(0,76){c)}
    \end{overpic}
    \hfill
    \begin{overpic}[width=0.45\textwidth]{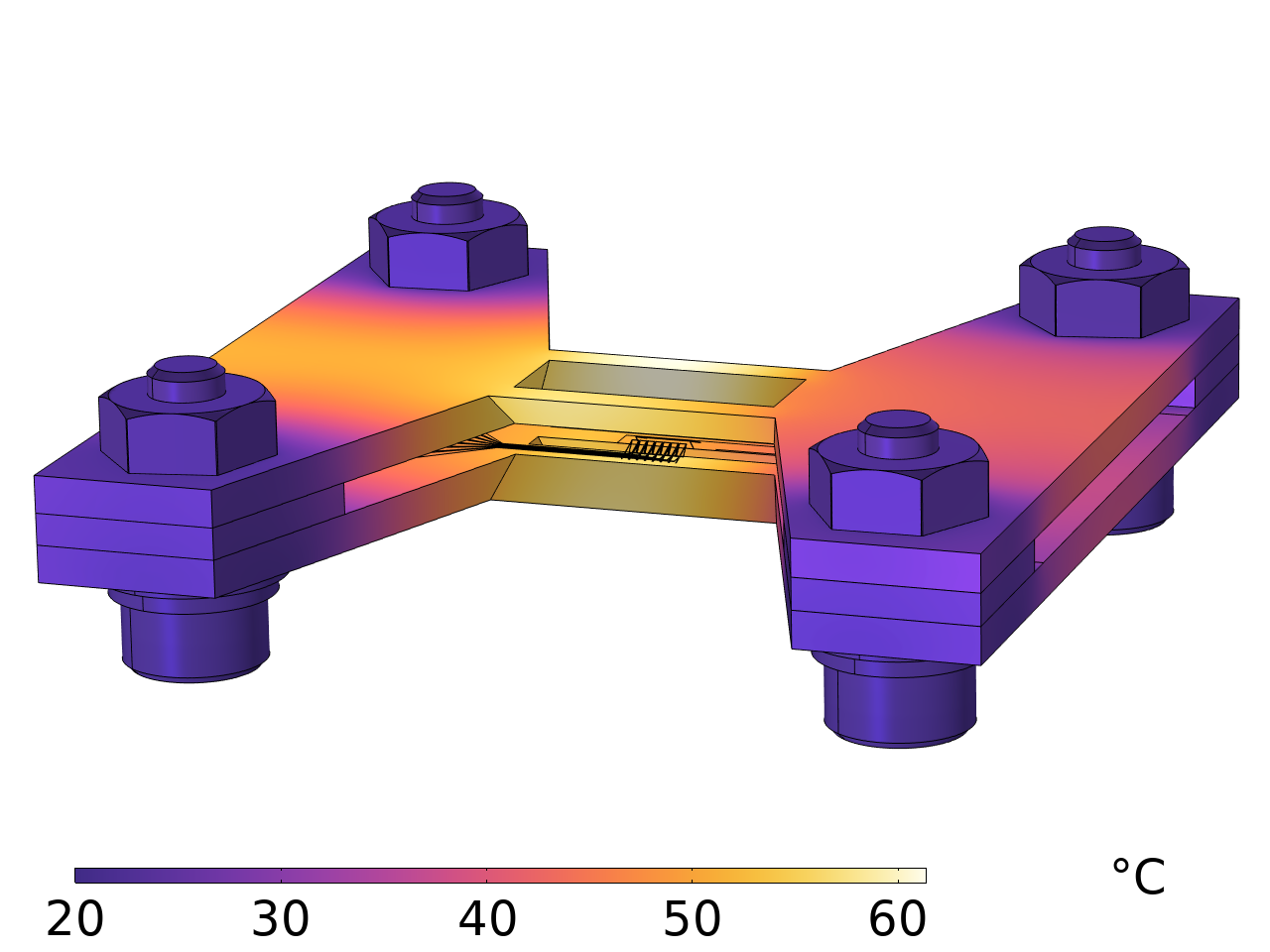}
        \put(0,76){d)}
    \end{overpic}

    \caption{
    \textbf{Schematic representations of the 3D-printed ion trap geometries analyzed in this work.}
    (a) Proposed 3D-printed ion trap with a skeleton electrode structure designed to reduce electric-field noise. The wire-frame electrodes have a diameter of 20~$\mu$m.
    (b) Side view of the 3D-printed trap showing a 400~$\mu$m distance between opposing electrodes, corresponding to an ion-to-electrode distance of 200~$\mu$m.
    (c) Zoom-in view of the segmented electrode “teeth” structure. Each electrode tooth is 170~$\mu$m wide, with a 9~$\mu$m gap between adjacent teeth.
    (d) Simulated steady-state temperature distribution of the trap under 600~V RF amplitude and 30~MHz drive frequency, showing a maximum temperature near 60~$^{\circ}$C.
    %In all subfigures, blue elements represent RF electrodes, and yellow elements represent DC electrodes.
    %The ion is located at the geometric center of the trap, equidistant from the closest electrode surfaces.
    %The same ion-to-electrode distance of 200 $\mu$m is maintained in both the blade and 3D-printed traps, allowing a direct and controlled comparison of electric-field-induced heating behavior between the two geometries.
    In all subfigures, blue elements denote RF electrodes and yellow elements denote DC electrodes. The ion is located at the geometric center of the trap, equidistant from the nearest electrode surfaces. 
    }
    \label{fig: traps}
    
\end{figure}

Figure \ref{fig: traps} illustrates the schematic diagrams and side views of the 3D-printed skeleton trap, highlighting the hollowed electrode structure and radial symmetry. For comparison, Figure \ref{fig: heating mechanism of the blade trap}(a) shows the schematic diagram of a conventional blade trap using the same ion-to-electrode distance.
Both trap designs maintain an ion-to-electrode spacing of 200~$\mu$m, defined as the shortest distance between the ion and the nearest point on any electrode surface. This design choice is motivated by practical fabrication constraints in blade traps: further miniaturization would require extremely thin and fragile endcap electrodes, potentially compromising mechanical robustness and alignment accuracy. 
Using the same distance in the 3D-printed trap ensures that any differences in performance, particularly heating rate, are attributable to trap geometry rather than differences in confinement strength.

The blade trap uses a conventional quadrupole arrangement of four opposing blade-shaped electrodes.
In contrast, the 3D-printed trap adopts a skeleton geometry that strategically removes bulk electrode material while preserving radial symmetry.

\subsection{RF Drive Parameters and Trap Characterization}
To simulate realistic ion confinement, three interdependent parameters must be chosen: 
the ion-to-electrode distance,
the RF voltage amplitude $V_{RF}$, 
and the RF drive frequency $\Omega_{RF}$.
These are constrained by trap stability conditions, material limits, and desired trap frequency.

\begin{figure}[t]
    \centering
    \begin{overpic}[width=0.5\textwidth]{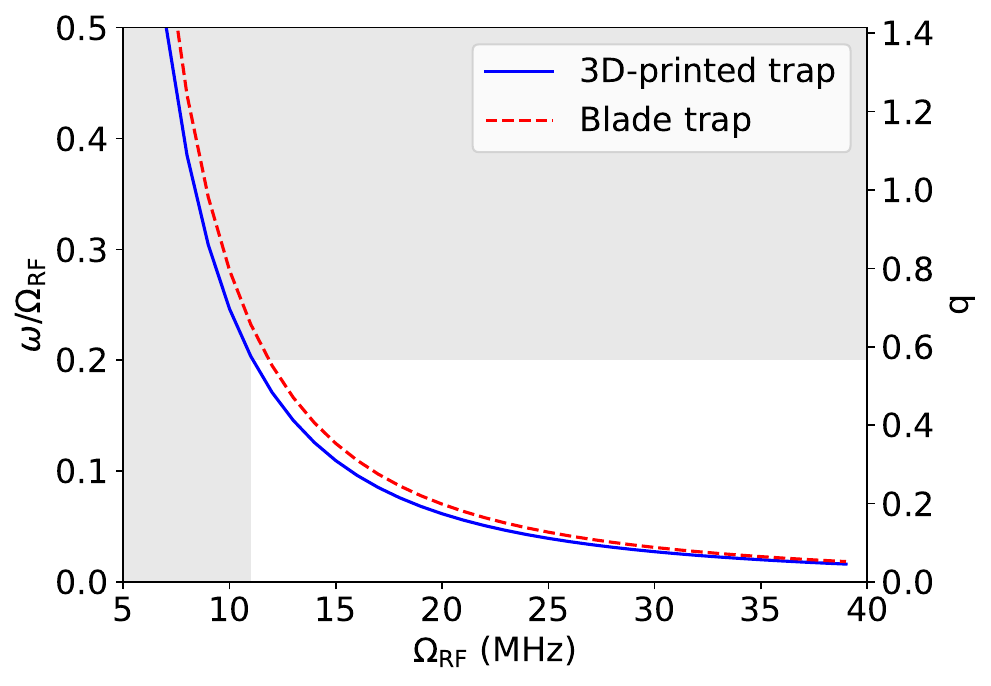}
        \put(-2,69){ a)}   % (x,y) in percent of width/height
    \end{overpic}
    \hfill
    \begin{overpic}[width=0.45\textwidth]{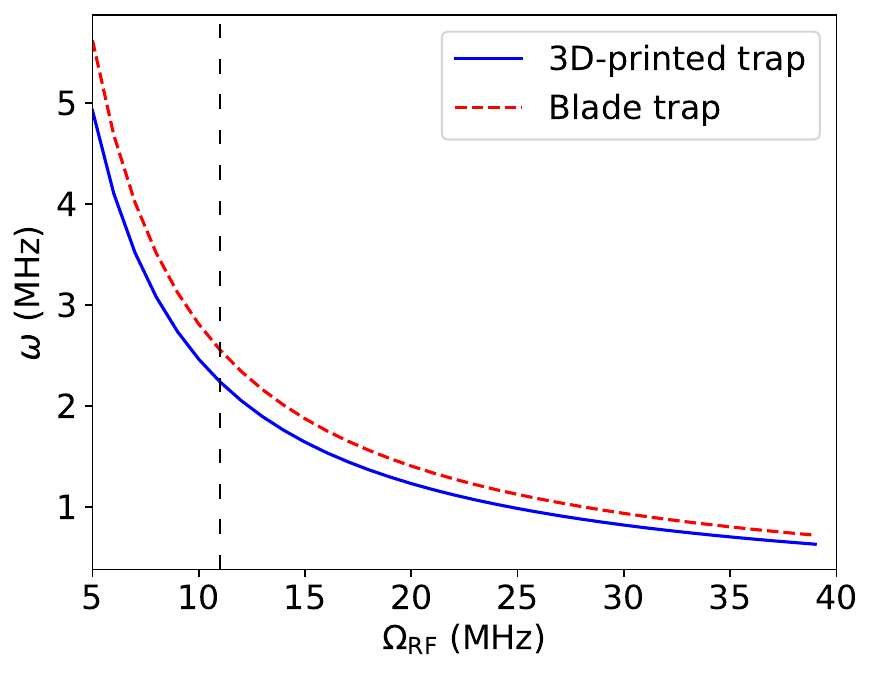}
        \put(0,76){b)}
    \end{overpic}
    \caption{
    \textbf{RF trap stability and secular frequency characteristics for the 3D-printed and blade traps.}
    (a) Ratio of the radial secular frequency $\omega$ to the RF drive frequency $\Omega_{RF}$ as a function of $\Omega_{RF}$,
    calculated for both trap geometries with a fixed RF voltage amplitude of 150~V.
    The grey-shaded region indicates unstable trapping conditions where the ratio exceeds 0.2, violating the commonly accepted stability criterion for linear RF traps.
    (b) Radial secular frequency $\omega$ as a function of RF drive frequency for a $^{171}$Yb$^+$ ion under the same RF voltage.
    The dashed vertical line at 11~MHz indicates the selected operating point, which yields a radial trap frequency of 2.24~MHz in the 3D-printed trap.
    This point lies near the upper boundary of the stable region and is chosen to maximize confinement strength while ensuring trap stability.
    }
    \label{fig: traps_freq}
\end{figure}

In this work, the RF voltage amplitude is fixed at 150~V, based on practical operating experience from existing ion trap systems \cite{Xu:2025aa}.
We then sweep the RF drive frequency to determine the stability region and achievable secular frequencies for a $^{171}$Yb$^+$ ion.
Figure \ref{fig: traps_freq}(a) shows the ratio $\omega / \Omega_{RF}$ as a function of $\Omega_{RF}$, where $\omega$ is the radial secular frequency. 
The shaded region indicates the unstable regime beyond the commonly accepted limit of $\omega / \Omega_{RF} > 0.2$ \cite{Kienzler2015}.
Figure \ref{fig: traps_freq}(b) shows the corresponding secular frequency as a function of drive frequency.
From these plots, we select an RF drive frequency of 11~MHz, yielding a radial trap frequency of 2.24~MHz, which lies near the upper stability boundary. This choice ensures strong confinement while remaining within stable operating parameters.

\subsection{Thermal and Structural Stability Modeling}
To assess the performance limits of the 3D-printed trap under high-power RF operation,
we conducted a comprehensive simulation study using COMSOL Multiphysics, incorporating both temperature rise analysis and mechanical eigenfrequency evaluation.

\subsubsection{Thermal Modeling of RF-Induced Heating.}

Ion trap heating originates primarily from two sources: Joule heating in the metallic electrodes and dielectric absorption in the insulating substrate. Heat is dissipated from the trap via conduction through mechanical mounting interfaces, as well as radiation from electrode surfaces into the surrounding environment~\cite{Dolezal:2015aa}.

Following a methodology adapted from Doležal et al.~\cite{Dolezal:2015aa}, we implement a two-step simulation process. First, we use the AC/DC module of COMSOL to solve for the electromagnetic field distribution within the trap geometry, calculating local power dissipation based on the applied voltage, frequency, and the material properties. In the second step, this dissipation profile is imported into the heat transfer module, where the resulting steady-state temperature map is computed by solving the heat conduction equation with the ambient temperature fixed at 20$^{\circ}$C. The simulation includes both conductive and radiative heat removal mechanisms.

The electrodes are modeled as gold, and the substrate is based on FR4, a commonly used glass-reinforced epoxy laminate. 
%Under high-voltage conditions (600~V RF amplitude and 30~MHz drive frequency), the total RF-induced power dissipation is approximately 182~mW. The resulting steady-state temperature rise leads to a maximum local temperature of approximately 60$^{\circ}$C, 
%as shown in the thermal simulation map in figure \ref{fig: traps}(d),
%indicating that the 3D-printed trap remains within safe thermal limits and is unlikely to undergo material degradation or structural deformation during normal operation.
Under high-voltage operating conditions of 600~V RF amplitude and 30~MHz drive frequency, the total RF-induced power dissipation is approximately 182~mW. While the RF voltage amplitude was fixed at 150~V for trap operation and characterization, consistent with practical conditions in many ion trap systems, we intentionally applied a much higher voltage amplitude of 600~V in the thermal simulation to assess the trap’s robustness under elevated electrical stress. This stress-test scenario demonstrates that even under extreme conditions, the 3D-printed structure remains thermally stable, reinforcing its suitability for scalable ion trap applications. 
The resulting steady-state temperature rise leads to a maximum local temperature of approximately 60$^{\circ}$C, indicating that the 3D-printed trap remains within safe thermal limits and is unlikely to undergo material degradation or structural deformation during normal operation. 
The simulated steady-state temperature distribution is shown in Figure~\ref{fig: traps}(d).

\subsubsection{Mechanical Resonance Analysis.}

In addition to thermal modeling, we carried out a structural eigenfrequency analysis using COMSOL's Structural Mechanics module. This step ensures that the trap does not exhibit mechanical resonances that could be excited by periodic RF driving or environmental vibrations.

The results show that the fundamental mechanical modes of the 3D-printed skeleton structure lie in the kilohertz range, far below the 30~MHz RF drive frequency. This large separation between RF and mechanical eigenfrequencies eliminates concerns about resonant amplification of structural vibrations, ensuring long-term mechanical stability and reducing susceptibility to vibration-induced electric field noise.

\subsection{Motivation for Skeleton Design}
The skeleton architecture of the 3D-printed trap significantly reduces the electrode surface area exposed to the ion, 
thereby mitigating electric field noise from nearby fluctuating patch potentials. This hollowed structure aims to suppress heating by minimizing the dominant noise contributions, 
which are known to arise primarily from electrode regions closest to the ion.

Additionally, the trap is constructed with periodic 
%"teeth" and
segmented trap electrodes with
narrow gaps (9~$\mu$m) between them. 
This 
periodic segmentation 
provides a useful design degree of freedom for shaping the spatial electric field environment and targeting noise suppression.

Later sections of this work explore how variations in electrode spacing affect heating behavior.
In particular, we investigate how repositioning the electrode edges away from regions of maximum heating contribution can lower the total heating rate, even if the overall electrode area is increased.
This informs the design of future optimized traps and demonstrates the potential of 3D printing for precise electrode engineering.

\section{Calculation Methodology}\label{sec: Calculation Methodology}

To compare the heating behavior of different ion trap geometries, we employ a theoretical framework that models the interaction between the trapped ion and fluctuating electric fields arising from electrode surface noise.
This section outlines the approach used to calculate the heating rates for both the blade and 3D-printed skeleton traps, including the treatment of patch potentials, electric field evaluation, and noise spectral density \cite{Low2011,Brownnutt2015}.

\subsection{Modeling Electric Field Noise from Patch Potentials}
The dominant source of motional heating in ion traps at room temperature is believed to originate from microscopic, 
%time-varying 
near-resonant RF
potential 
%fluctuations 
noise
on the electrode surfaces, known as patch potentials \cite{Turchette2000,Labaziewicz2008,Safavi-Naini2011}. 
These fluctuations generate spatially and temporally 
%varying
fluctuating
electric fields at the ion’s location.

We consider that the ion is confined in a harmonic potential and initially occupies the quantum mechanical ground state of motion. 
The electric field noise $\mathbf{E}(t)$ at the ion location can drive transitions to excited motional states.
For the $k$-th motional direction (e.g. radial or axial), 
we denote the corresponding field component as $\mathbf{E}_k(t)$, 
and evaluate the heating rate using first-order time-dependent perturbation theory, as shown in equation \eqref{eq: transition rate from 0 to 1}.
 
\subsection{Evaluation of Electric Field from Patch Potentials}

To evaluate $E_k(\mathbf{r},t)$ in equation \eqref{eq: the power spectral density of the fluctuating electric field}, we consider the field at the ion generated by a fluctuating potential $\phi(\mathbf{r}',t)$ located on the electrode surface $\sigma$.
The field component in direction $k$ at the ion position $\mathbf{r}$ is:
\begin{equation}
    E_k (\mathbf{r},t) =
    - \int_\sigma \phi (\mathbf{r}',t) 
    \frac{\partial}{\partial n'} \nabla_k G(\mathbf{r},\mathbf{r}')
    \, d\mathbf{r}'.
\end{equation}
Here, $G(\mathbf{r},\mathbf{r}')$ is the Green’s function of Laplace’s equation that satisfies Dirichlet boundary conditions for the trap geometry,
and $\partial / \partial n'$ denotes the derivative normal to the electrode surface at $\mathbf{r}'$ \cite{Low2011}.
To determine the total field noise at the ion, we compute the spatial integral over the electrode surface and evaluate the temporal correlation between surface points. 
The spectral density becomes
%\begin{align} \label{eq: total heating rate}
 %   S_{E_k} = & 2 \int_{\sigma''} \int_{\sigma'}
 %   \mathbf{F}[\langle \phi(\mathbf{r}',t) \phi(\mathbf{r}'',t+\tau) \rangle] \times \notag \\
  %  &\left [\frac{\partial}{\partial n'} \nabla_k G(\mathbf{r},\mathbf{r}')
  %  \cdot 
  %  \frac{\partial}{\partial n''} \nabla_k G(\mathbf{r},\mathbf{r}'')
  % \right ] 
  %  \, d \mathbf{r}' d\mathbf{r}''.
%\end{align}
\begin{equation} \label{eq: total heating rate}
    S_{E_k} =  2 \int_{\sigma''} \int_{\sigma'}
    \mathbf{F}[\langle \phi(\mathbf{r}',t) \phi(\mathbf{r}'',t+\tau) \rangle] 
    \left [\frac{\partial}{\partial n'} \nabla_k G(\mathbf{r},\mathbf{r}')
    \cdot 
    \frac{\partial}{\partial n''} \nabla_k G(\mathbf{r},\mathbf{r}'')
    \right ] 
    \, d \mathbf{r}' d\mathbf{r}''.
\end{equation}
This formalism enables a spatially resolved heating rate analysis, allowing us to identify dominant noise contributors and assess the influence of geometry on total heating.

\subsection{Patch-by-Patch Heating Rate Analysis}

To elucidate the spatial origin of ion heating in complex electrode geometries, we adopt a patch-resolved analysis approach, in which the electrode surface is discretized into a large number of small surface elements, or “patches.”
Each patch is assumed to exhibit an independent fluctuating potential, characterized by a localized time-varying surface voltage.
By modeling each patch as an isolated noise source, we are able to calculate its individual contribution to the heating rate of the trapped ion.

For each patch, we compute the electric field it generates at the ion’s position and project this field along the direction of the motional mode of interest, whether radial or axial.
Using first-order perturbation theory, we then evaluate the associated heating rate by determining the power spectral density of the electric field at the ion’s secular frequency.
Summing the contributions from all patches provides an estimate of the total heating rate. 
This procedure not only yields a global measure of the ion’s susceptibility to electric field noise but also reveals how different surface regions contribute to heating.

Importantly, this methodology enables a detailed spatial map of heating sources on the electrode surfaces, offering valuable insight into mode-dependent behaviors.
For example, it becomes possible to identify whether certain regions of the trap contribute predominantly to axial heating, while others play a larger role in radial heating. 
Such information is critical for understanding how trap geometry influences noise coupling and for guiding future design strategies aimed at suppressing heating through geometric optimization.

Numerical implementation of this analysis is carried out using finite-element solvers, specifically COMSOL Multiphysics, to compute the electrostatic Green’s function $G(\mathbf{r},\mathbf{r}')$ and its spatial derivatives with high precision.
These quantities are then processed with custom-developed code to evaluate the temporal correlation functions of the patch potentials and extract the corresponding electric field spectral densities.
The resulting data form the basis for the mode-resolved and spatially resolved heating rate distributions discussed in the subsequent sections of this work.

\section{Results and Analysis}\label{sec5}

In this section, we present a comparative study of the motional heating rates in the conventional blade trap and the proposed 3D-printed skeleton trap. The analysis includes total heating rate evaluation, spatial mapping of noise contributions, mode-dependent behavior, and the effects of geometric optimization.

\subsection{Total Heating Rate Comparison}

Using the patch potential model described in section \ref{sec: Calculation Methodology},
we compute the total heating rates for both trap geometries under identical confinement conditions.
Both traps are designed with an ion-to-electrode distance of 200~$\mu$m and operated with an RF voltage amplitude of 150~V and an RF drive frequency of 11~MHz,
yielding a radial secular frequency of 2.24~MHz for $^{171}$Yb$^+$.

The simulation reveals that the total heating rate in the 3D-printed trap is reduced by over 50\% compared to that of the blade trap.
This significant reduction is attributed to the skeleton structure, which minimizes the nearby electrode surface area and thereby suppresses the influence of noisy patches close to the ion.

\subsection{ Spatial Distribution of Patch-Induced Heating}

\begin{figure}[t]
    \centering
    % Top row
    \begin{overpic}[width=0.5\textwidth]{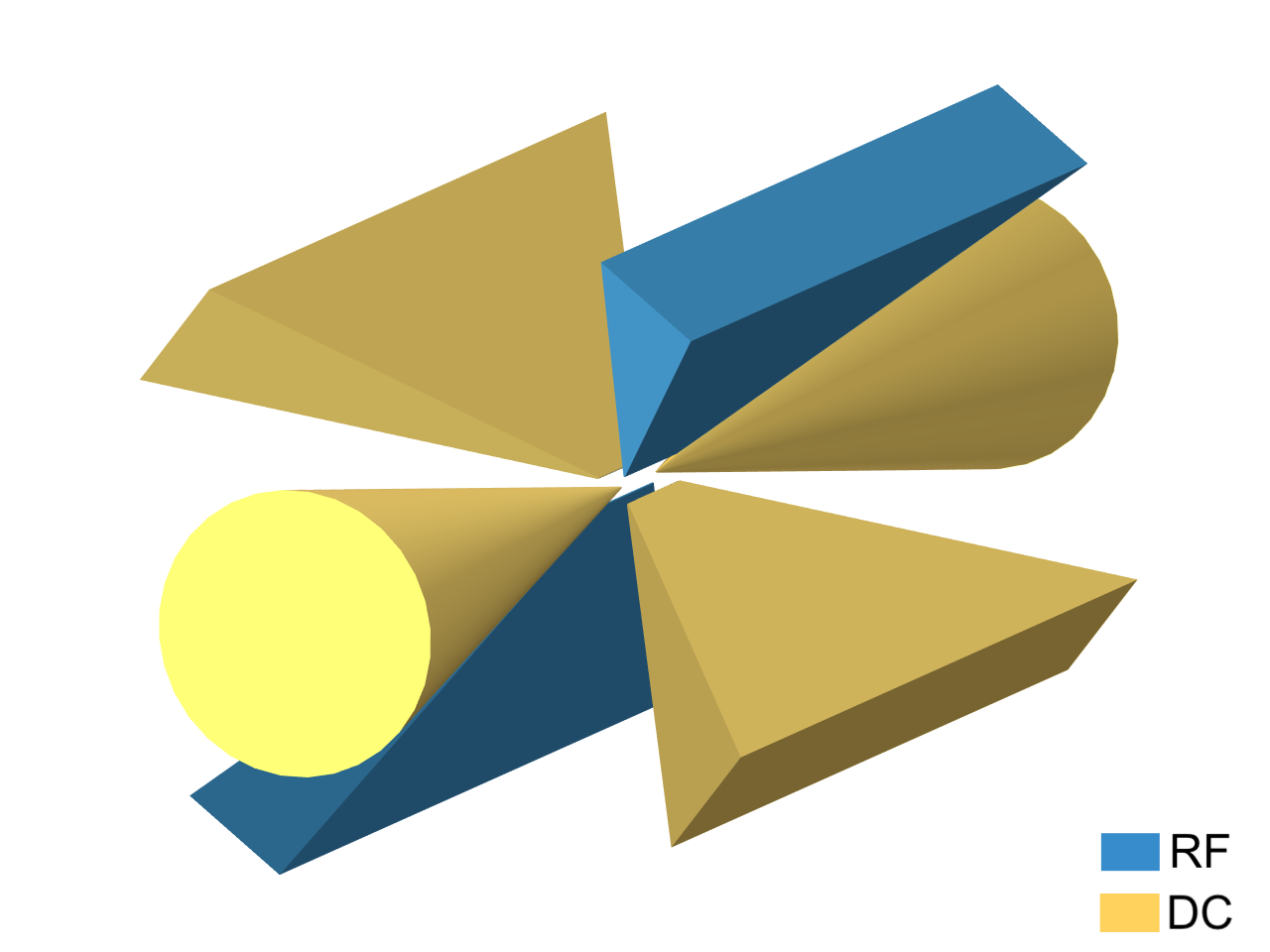}
        \put(0,76){a)}
    \end{overpic}

    \vspace{0.5cm} % space between rows
    \begin{overpic}[width=0.45\textwidth]{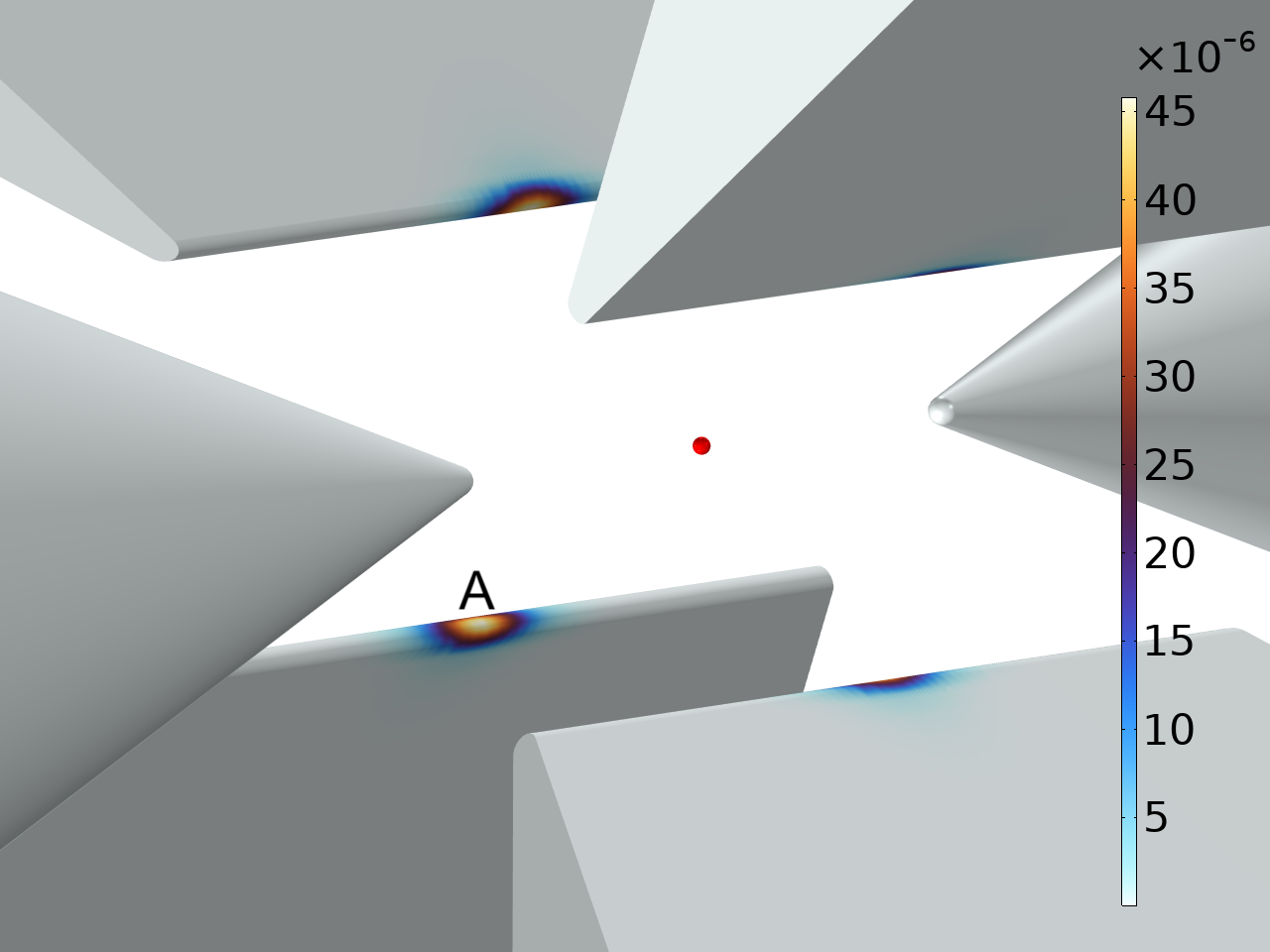}
        \put(0,76){ b)}   % (x,y) in percent of width/height
    \end{overpic}
    \hfill
    \begin{overpic}[width=0.45\textwidth]{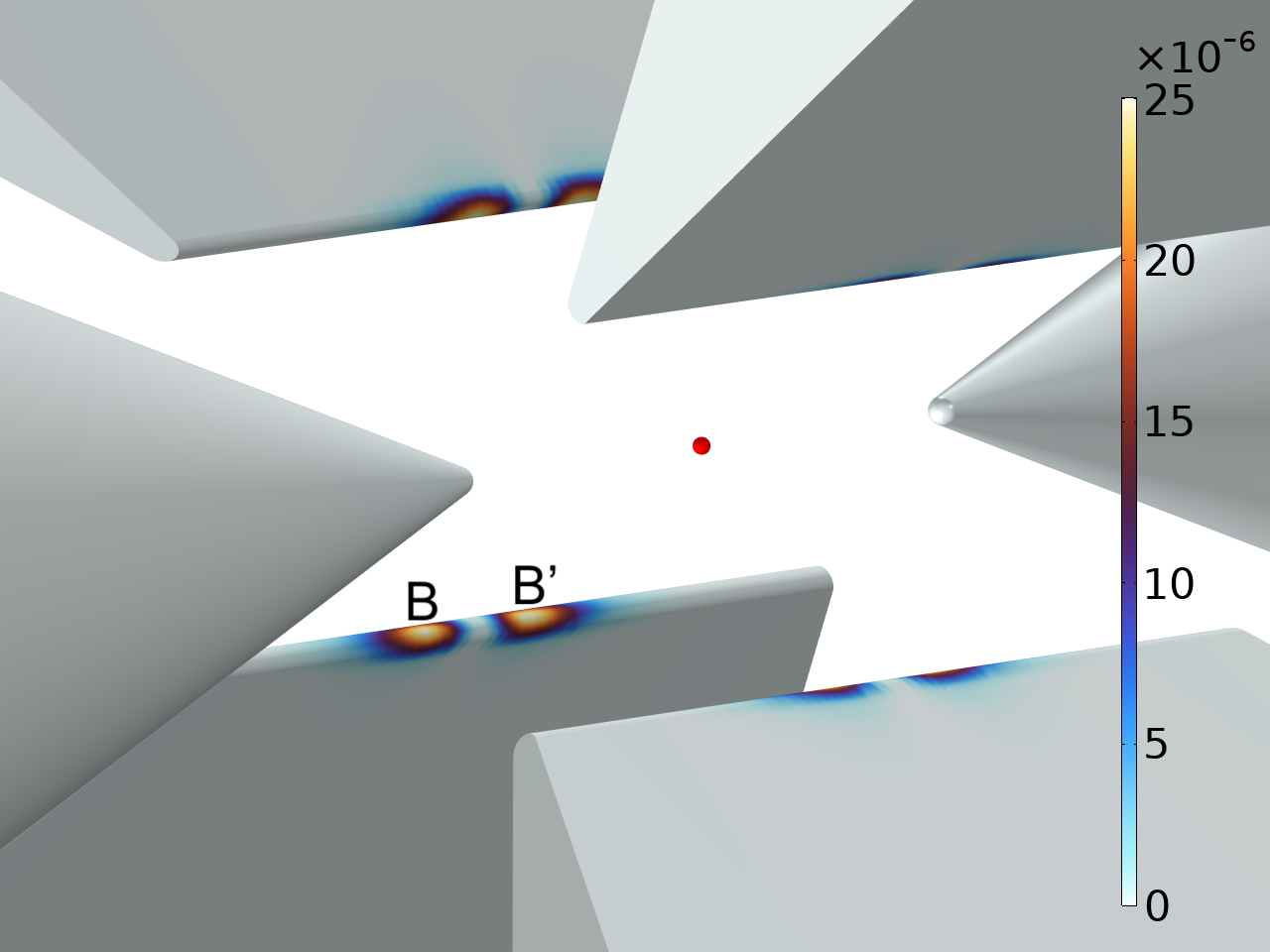}
        \put(0,76){c)}
    \end{overpic}
    \caption{
    \textbf{Schematic and spatial heating analysis of the conventional blade ion trap.} 
    (a)~Schematic representation of the conventional blade trap used for trapped‑ion confinement. Blue electrodes correspond to RF potentials and yellow electrodes correspond to DC potentials. The ion is located at the geometric center of the quadrupole formed by four blade electrodes.  
    The same 200~$\mu$m ion-to-electrode distance is used for both the skeleton trap and the blade trap to enable a controlled comparison of electric-field-induced heating between the two designs.
    (b)~Spatial distribution of heating‑rate contributions to the \textbf{radial} motional mode from localized patch potentials on the electrode surfaces. Each color map illustrates the normalized contribution of individual surface patches to the total heating rate, highlighting dominant noise regions near the ion. Point A marks the location of the patch that gives the highest heating rate in the radial motional mode, corresponding to the region closest to the ion (200~$\mu$m). The red sphere represents the ion.
    (c)~Spatial distribution of heating‑rate contributions to the \textbf{axial} motional mode. In contrast to the radial case, axial‑mode heating exhibits a non‑monotonic profile, with maximum contributions arising from patches located at intermediate distances (approximately 110~$\mu$m from the ion) indicated by points B and B'. The red sphere again denotes the ion position.
    }
    \label{fig: heating mechanism of the blade trap}
\end{figure}

\begin{figure}[t]
    \centering
    \begin{overpic}[width=0.49\textwidth]{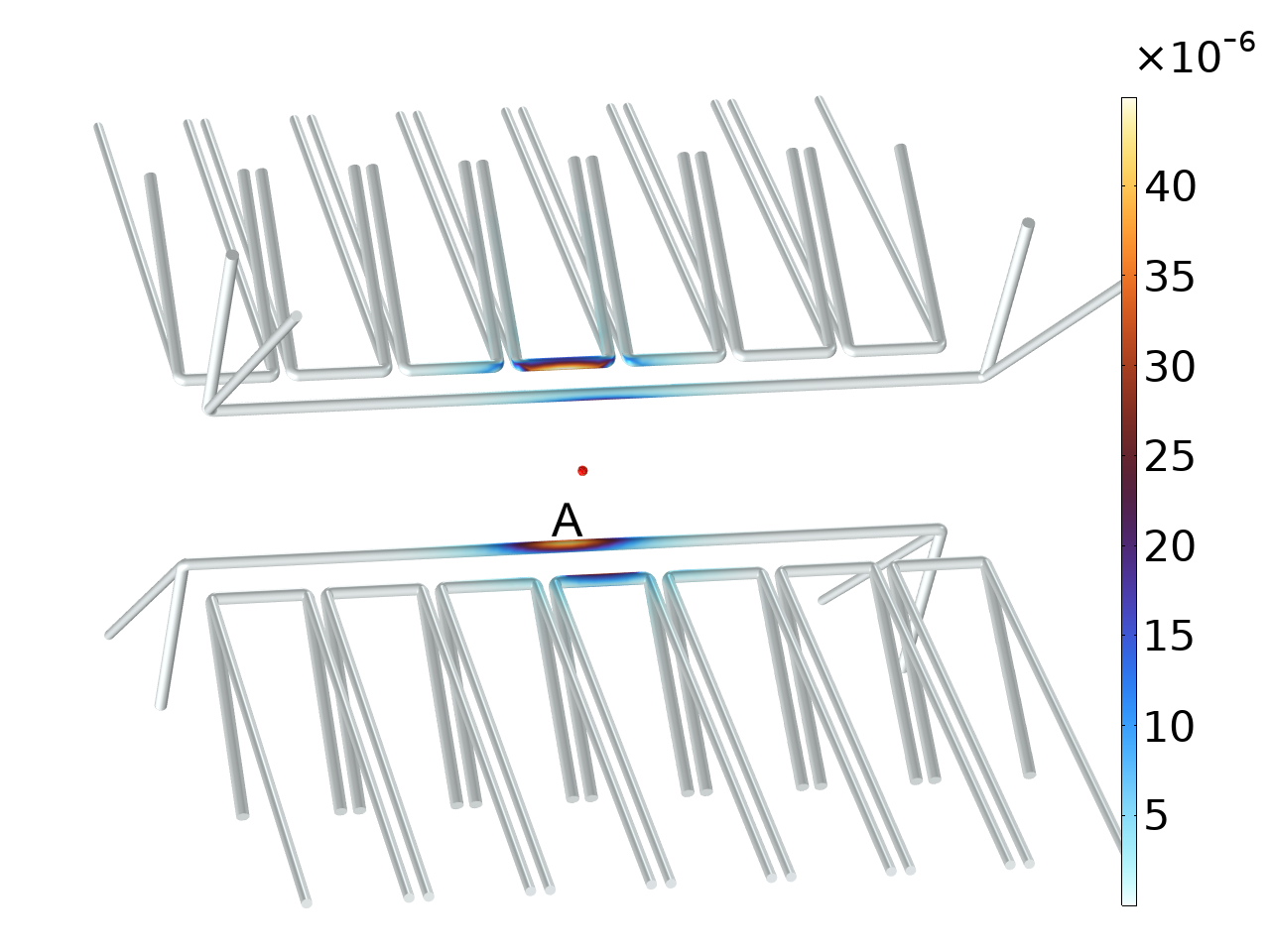}
        \put(0,76){ a)}   % (x,y) in percent of width/height
    \end{overpic}
    \hfill
    \begin{overpic}[width=0.49\textwidth]{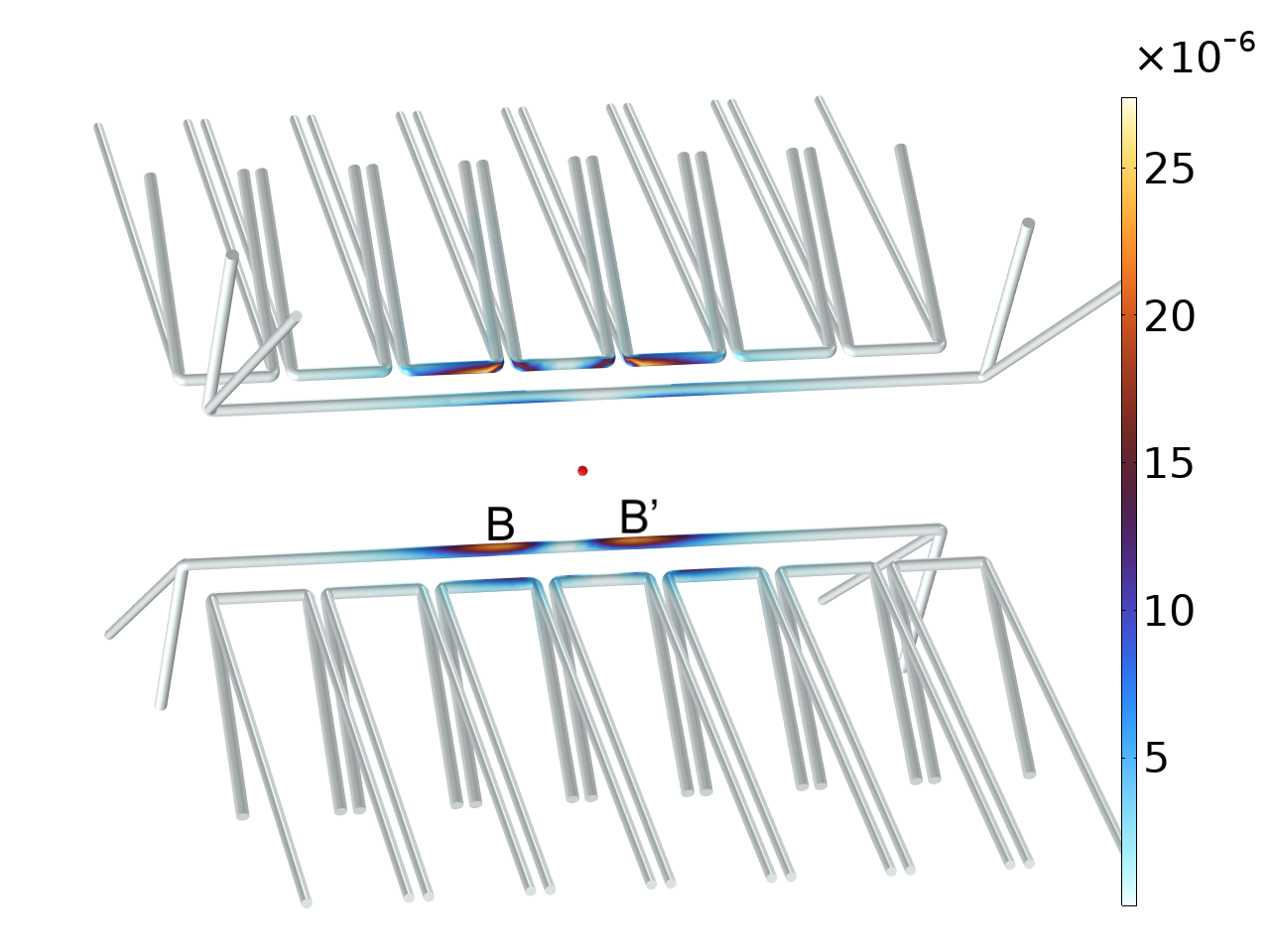}
        \put(0,76){b)}
    \end{overpic}
    \caption{
    \textbf{Spatially resolved analysis of electric‑field‑induced heating in the 3D‑printed skeleton ion trap.} 
    (a)~Heating‑rate contributions to the \textbf{radial} motional mode from surface patches on the skeleton electrodes. 
    Point A marks the surface patch giving the highest radial‑mode heating rate, corresponding to the region closest to the ion (200~$\mu$m). The red sphere represents the ion. 
    (b)~Heating‑rate contributions to the \textbf{axial} motional mode. The red sphere again denotes the ion position. Each color map in (a) and (b) shows the relative contribution of individual surface patches to the total heating rate, where color intensity represents the normalized value as a fraction of the total. 
    Points B and B' indicate the surface patches giving the highest axial‑mode heating rates.
    These results highlight the spatial dependence of motional heating in the 3D‑printed skeleton trap, revealing mode‑specific regions of dominant electric‑field noise.
    Compared to the blade trap, the 3D-printed skeleton trap exhibits a substantial suppression of heating contributions, especially near the ion, due to its hollow, minimal-surface-area electrode structure.}
    \label{fig: heating mechanism of the 3D-printed trap}
\end{figure}

To elucidate the origin of the heating rate difference, we perform a spatially resolved analysis of patch contributions to the total heating rate.
Figures \ref{fig: heating mechanism of the blade trap}(b) and \ref{fig: heating mechanism of the blade trap}(c) show the heating rate contributions from different surface patches for the radial and axial modes in the blade trap, while figures \ref{fig: heating mechanism of the 3D-printed trap}(a) and \ref{fig: heating mechanism of the 3D-printed trap}(b) illustrate the same for the 3D-printed skeleton trap.

\begin{figure}[t]
    \centering
    \begin{overpic}[width=0.45\textwidth]{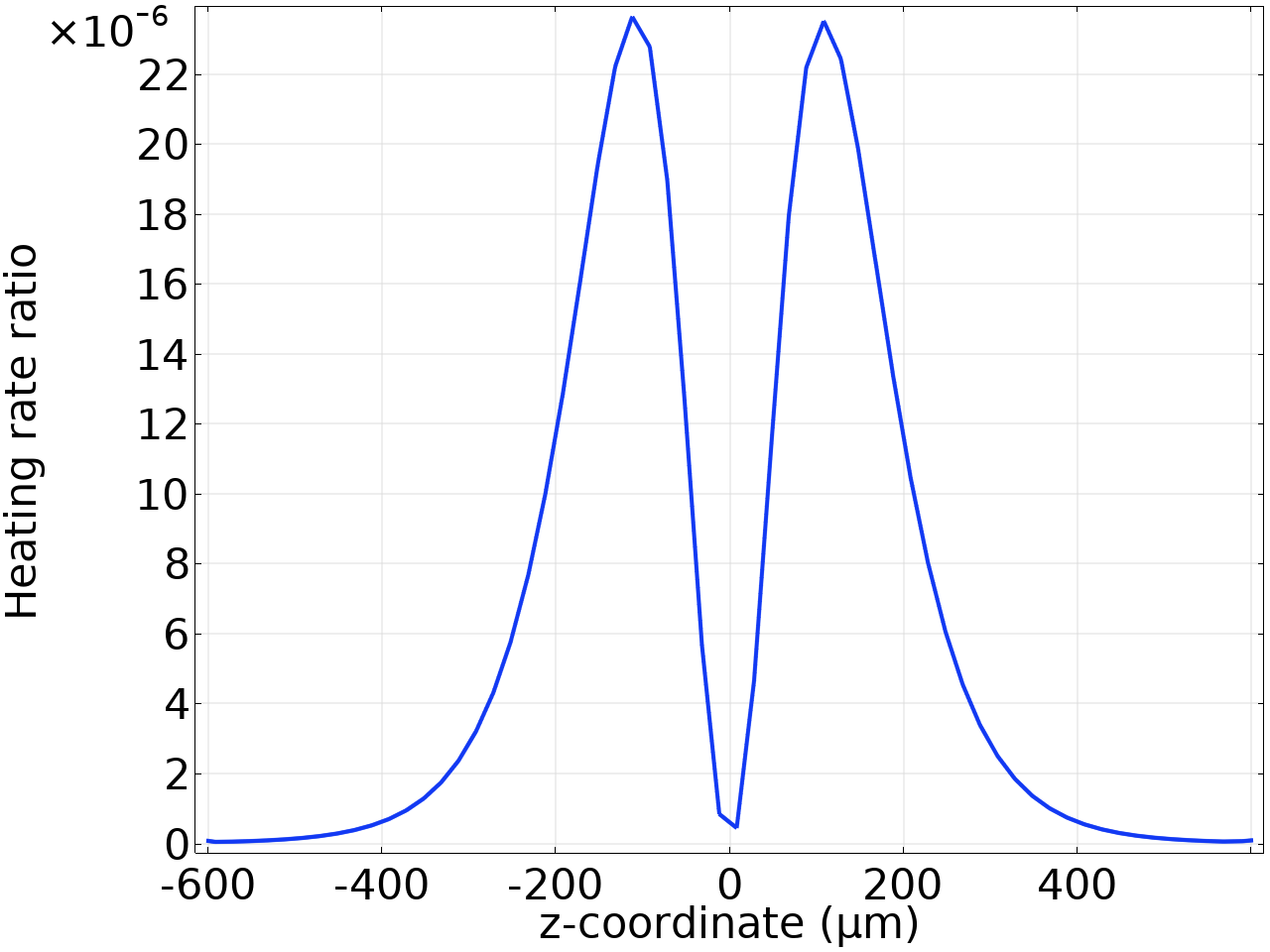}
        \put(0,76){ a)}   % (x,y) in percent of width/height
    \end{overpic}
    \hfill
    \begin{overpic}[width=0.45\textwidth]{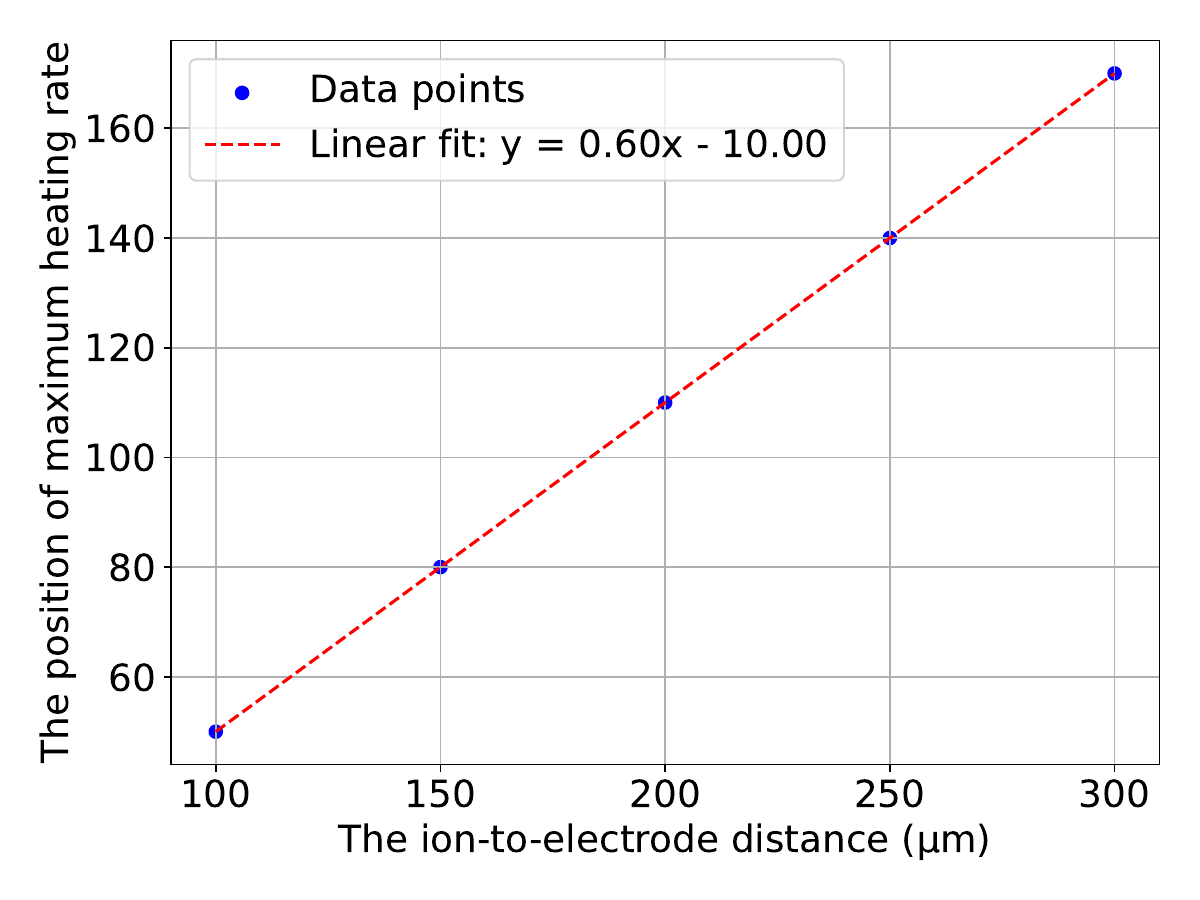}
        \put(0,76){b)}
    \end{overpic}
    \caption{
    \textbf{Axial-mode heating behavior in linear ion traps analyzed through patch potential simulations.}
    (a) Axial heating rate contributions from surface patch potentials along a single RF electrode. A distinct peak is observed at approximately 110~$\mu$m away from the ion position, illustrating that the dominant contribution to axial heating arises from patches not closest to the ion.
(b) Linear relationship between the ion-to-electrode distance and the axial position of maximum heating rate. Each data point represents a different trap geometry. The observed linear trend confirms that the location of dominant axial-mode heating shifts proportionally with trap size. The fitted line follows the equation $y = 0.6x - 10$, where $x$ is the ion-to-electrode distance and $y$ is the distance from the ion to the peak axial heating source.
    }
    \label{fig. heating rate patch in 3D-printed trap}
\end{figure}

We find that in both trap geometries, over 99\% of the total heating rate originates from surface patches located within 500~$\mu$m of the ion.
The 3D-printed skeleton trap’s hollow structure eliminates a large fraction of these contributing patches by design, resulting in the observed reduction in heating.

\subsection{Mode-Dependent Behavior and Field Directionality}

Our analysis reveals that the spatial distribution of heating contributions differs significantly between the radial and axial motional modes of the trapped ion. 
These differences stem primarily from the directionality of the electric field vectors generated by the fluctuating patch potentials on the electrode surfaces, as well as their alignment with the corresponding mode axes.

For the radial motional mode, the dominant contributions to the heating rate arise from electrode patches located nearest to the ion. This behavior is intuitive: the electric fields produced by nearby surface patches are both stronger in magnitude and more closely aligned with the radial direction of confinement.
As a result, these fields couple efficiently to radial motion, making these proximal patches the primary sources of radial heating.
This trend holds across both the conventional blade trap and the 3D-printed skeleton trap, indicating a general behavior rooted in near-field interactions.

In contrast, the heating profile for the axial motional mode displays a more complex spatial dependence. 
Notably, the maximum heating rate in this direction does not originate from the patches closest to the ion.
Instead, 
%in the 3D-printed trap, 
we observe that the peak contribution arises from patches located approximately 110~$\mu$m away from the ion in the axial direction. 
This non-monotonic behavior is attributed to the orientation of the electric fields generated by the surface patches. 
Patches situated very close to the ion produce fields that are predominantly radial in direction, 
contributing minimally to axial motion. 
As the distance from the ion increases, however, the geometry of the trap leads to an increasing axial component in the field vectors, which enhances the coupling to the axial mode and results in a localized peak in the heating rate.

\begin{figure}[t]
    \centering
    \begin{overpic}[width=0.45\textwidth]{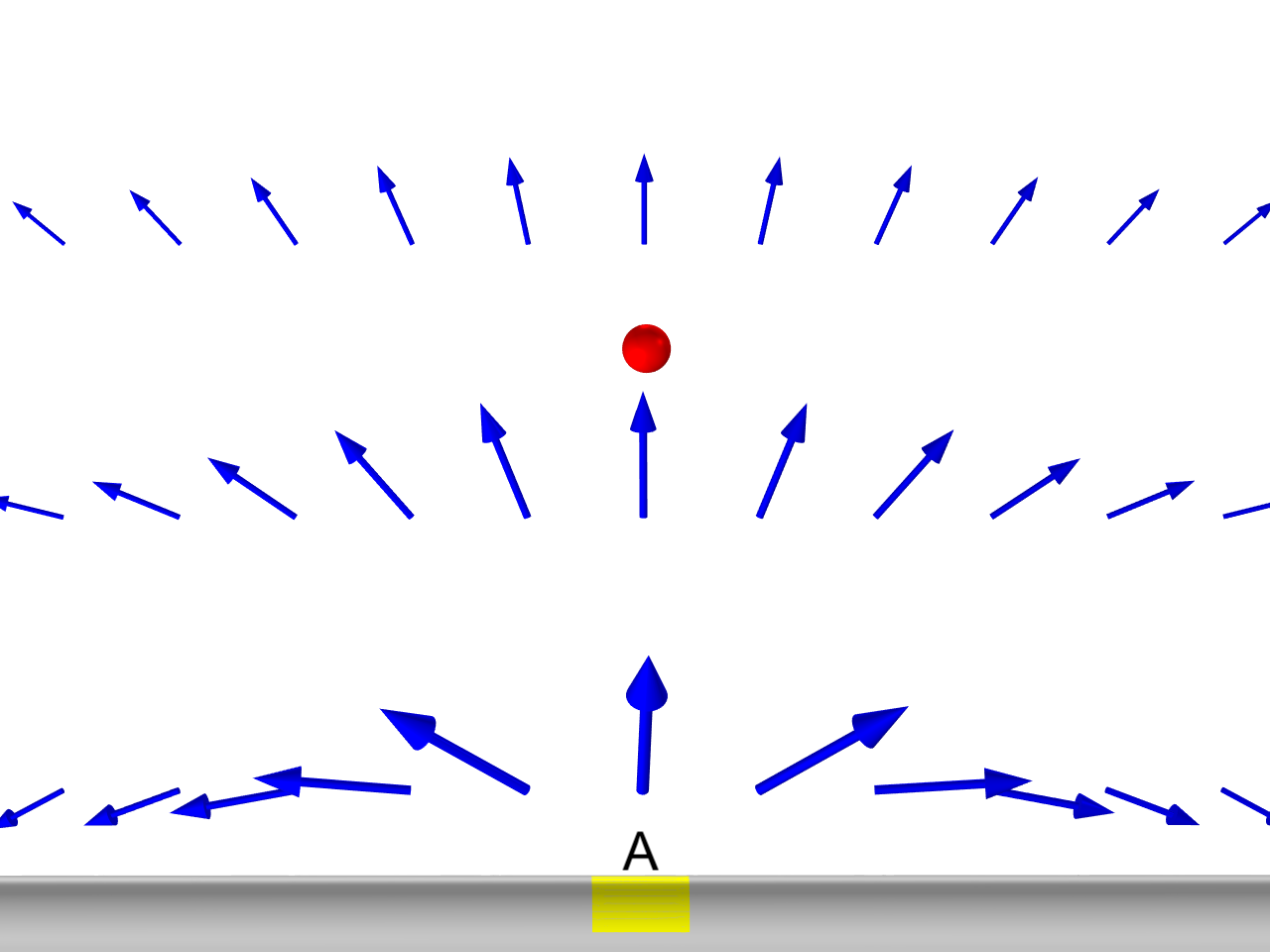}
        \put(0,76){ a)}   % (x,y) in percent of width/height
    \end{overpic}
    \hfill
    \begin{overpic}[width=0.45\textwidth]{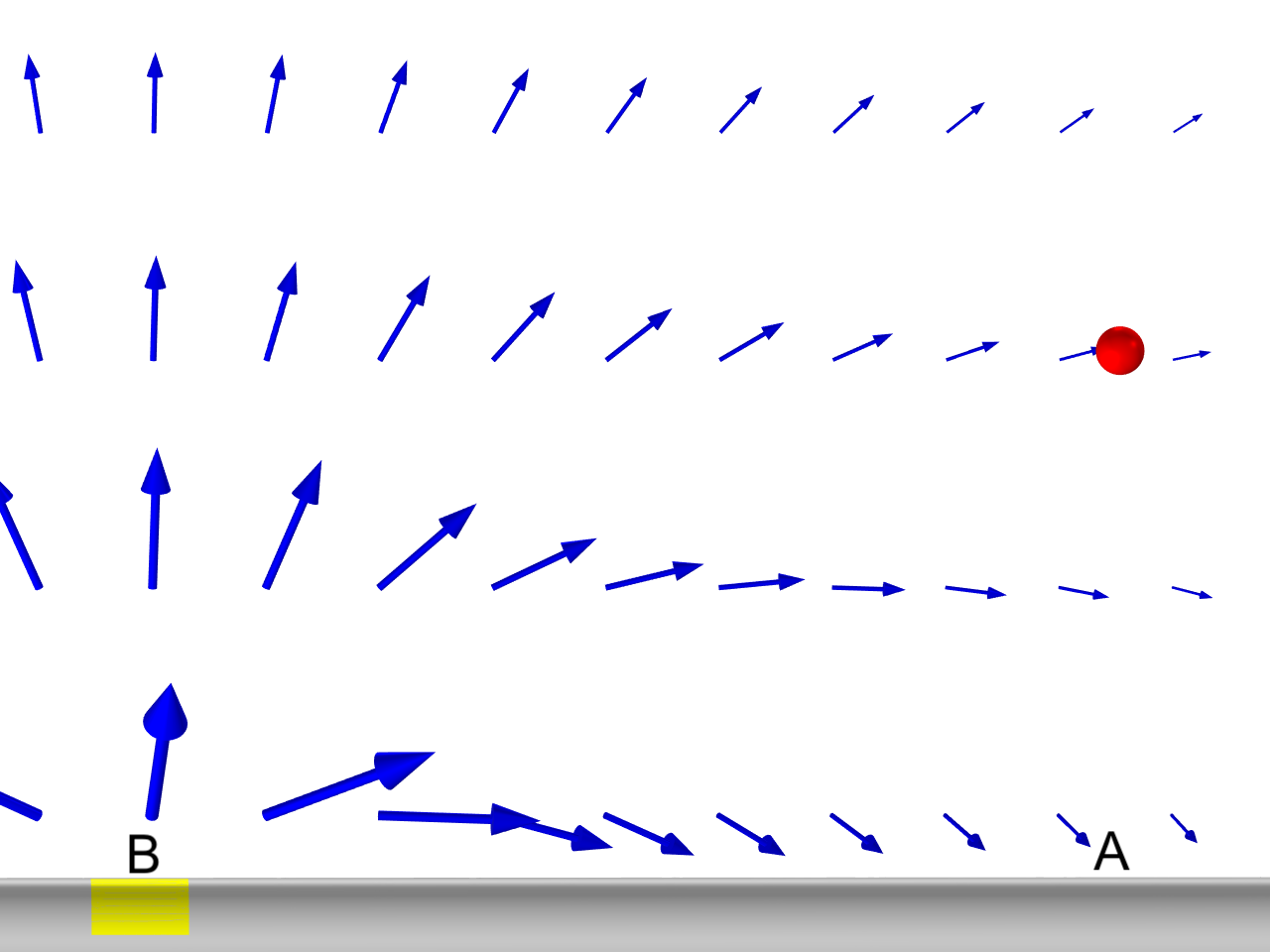}
        \put(0,76){b)}
    \end{overpic}
    \caption{
    \textbf{Electric‑field vector distributions illustrating the physical origin of mode‑dependent heating in both blade and 3D‑printed skeleton traps.} 
    (a)~Electric‑field vector distribution at the ion’s position generated by a surface patch located very close to the ion. The resulting field is predominantly radial with negligible axial component, contributing mainly to radial‑mode heating as shown in Figures \ref{fig: heating mechanism of the blade trap}(b) and \ref{fig: heating mechanism of the 3D-printed trap}(a). Point A corresponds to the same patch identified in those figures, marking the region that yields the highest radial‑mode heating rate at an ion‑to‑surface distance of 200~$\mu$m.  
    (b)~Electric‑field vector distribution from a patch located farther away, at a distance of approximately 110~$\mu$m from the ion. At this position, the axial field component becomes more pronounced, increasing its contribution to axial‑mode heating, consistent with the peaks observed in Figures \ref{fig: heating mechanism of the blade trap}(c) and \ref{fig: heating mechanism of the 3D-printed trap}(b). 
    Point B marks the patch responsible for these maximum axial‑mode contributions. 
    In both subfigures, the yellow patch marks the location of the localized fluctuating potential, the blue arrows represent the resulting electric‑field vectors, and the red dot indicates the ion position.  
    These visualizations provide a clear physical picture of how field directionality determines the spatially dependent heating behavior in trapped‑ion systems.
    }
    \label{fig: electric-field vector}
\end{figure}

This interpretation is supported by electric field vector visualizations, 
which illustrate the transition from radial-dominated to axial-contributing field components as the patch distance increases. 
%Figures \ref{fig. E-field patch in blade trap} presents representative vector field plots for both the blade trap and the 3D-printed trap, 
%highlighting how specific spatial regions contribute differently to axial and radial heating. 
%Representative vector field plots for the blade trap are shown in figures \ref{fig: heating mechanism of the blade trap}(d) and \ref{fig: heating mechanism of the blade trap}(e),
%while corresponding visualizations for the 3D-printed skeleton trap are presented in figures \ref{fig: heating mechanism of the 3D-printed trap}(d) and \ref{fig: heating mechanism of the 3D-printed trap}(e).
Representative vector field plots are shown in figures \ref{fig: electric-field vector}(a) and \ref{fig: electric-field vector}(b).
These figures highlight how specific spatial regions contribute differently to axial and radial heating.
These observations underscore the importance of not only minimizing overall surface area near the ion but also strategically shaping the electrode geometry to mitigate axial-mode heating by suppressing axial electric field components in critical regions.

%\begin{figure} [h] 
 
%    \caption{Electric field distribution generated by patch potentials on the 3D-printed trap.
%(a) Electric field from a patch potential located close to the ion, showing primarily radial directionality.
%(b) Electric field from a patch further from the ion, exhibiting an increased axial component similar to the behavior observed in the blade trap.
%These results support the conclusion that the directionality of electric field noise plays a key role in determining the spatial profile of axial mode heating, as seen in Figure \ref{fig. heating rate patch in 3D-printed trap} (b).}
%    \label{fig. E-field patch in 3D-printed trap}
%\end{figure}

\subsection{Electrode Optimization to Suppress Axial Heating}
Building on the spatial analysis of axial-mode heating contributions, 
we explore a strategy to reduce ion heating by geometrically optimizing the electrode layout. 
Our patch-resolved heating rate analysis revealed that the most significant contributions to axial heating originate from a localized region approximately 110~$\mu$m away from the ion. Motivated by this, we propose a modification to the 3D-printed skeleton trap design that realigns the gap between adjacent electrode “teeth” with this heating hotspot. The rationale is that by removing electrode material precisely at these critical positions, we can suppress the axial electric field components that drive heating in this mode.

To implement this strategy, we increased the width of each electrode tooth and repositioned the inter-electrode gaps so they coincide with the regions of peak heating.
%Despite the counterintuitive increase in total electrode surface area—due to the enlarged electrode segments—the optimization results in a net decrease in total axial heating. 
Despite the counterintuitive increase in total electrode surface area resulting from the enlarged electrode segments, the optimization results in a net decrease in total axial heating.
This is a significant observation: it demonstrates that axial heating can be reduced not merely by minimizing overall surface area, 
but by selectively shaping the geometry to suppress specific field components. 
In the optimized configuration, the regions that previously contributed most strongly to axial-mode heating no longer support electrode material, 
effectively eliminating their noise contribution.

\begin{figure}[t]
    \centering
    \begin{overpic}[width=0.5\textwidth]{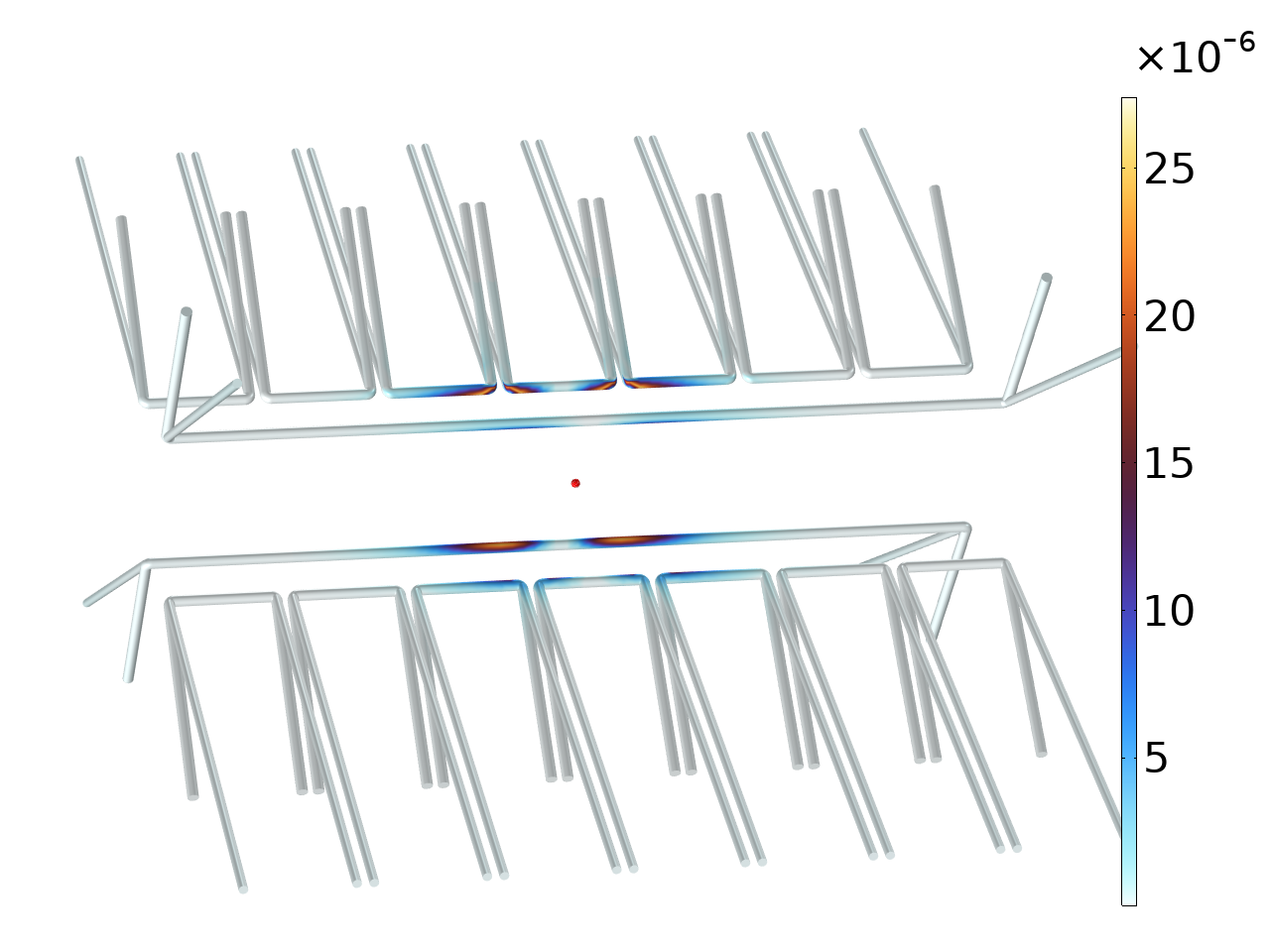}
        \put(0,68){ a)}   % (x,y) in percent of width/height
    \end{overpic}
    \hfill
    \begin{overpic}[width=0.45\textwidth]{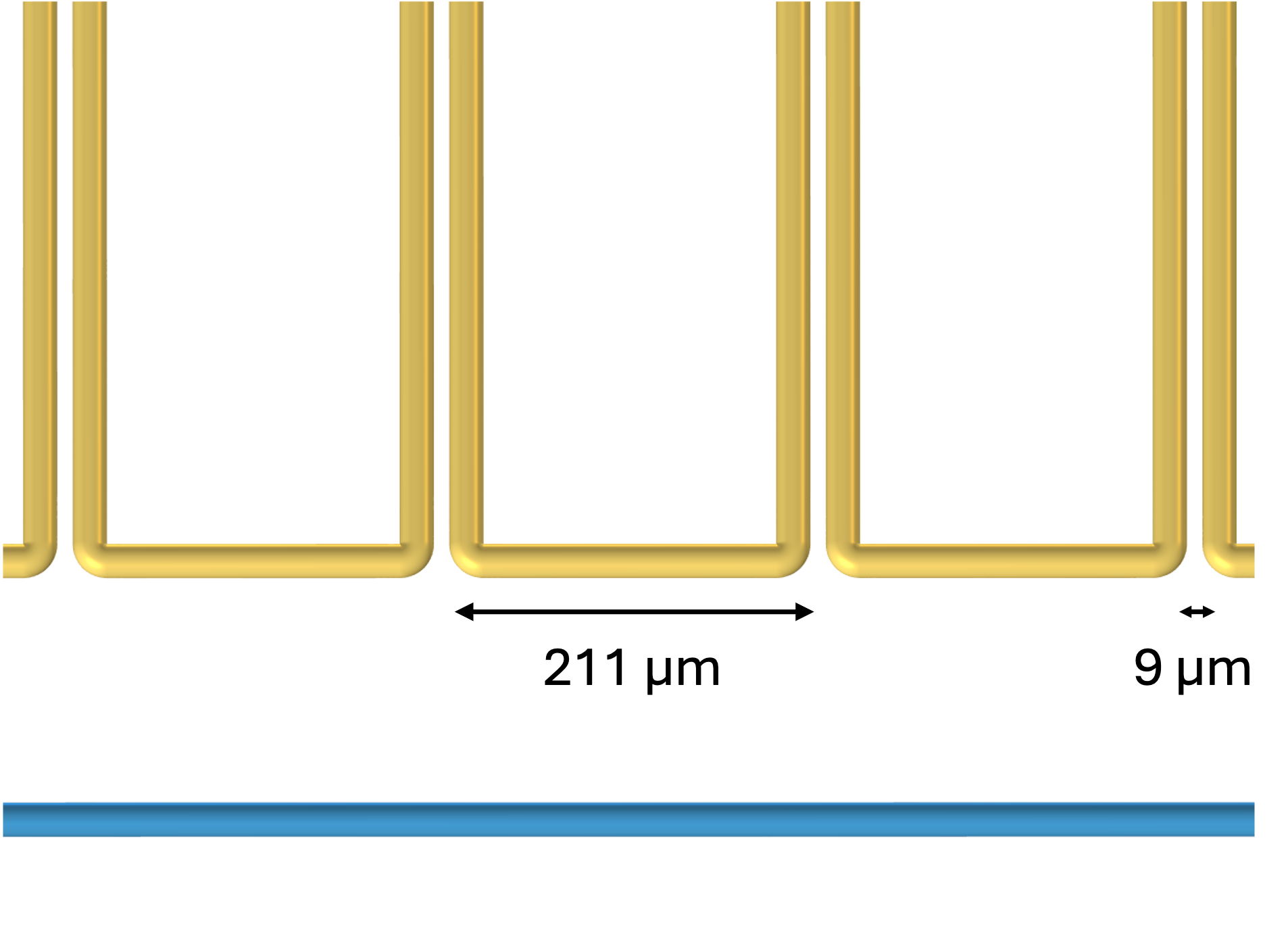}
        \put(0,76){b)}
    \end{overpic}
    \caption{
    \textbf{Spatial heating distribution and geometry of the optimized 3D-printed trap.}
    (a) Spatial distribution of heating rate contributions to the axial motional mode from surface patch potentials in the optimized 3D-printed skeleton trap. The electrode gaps have been repositioned to coincide with regions of peak axial heating observed in the original geometry, thereby suppressing dominant contributions.
    (b) Electrode geometry of the optimized trap.
    The width of each electrode tooth is increased to 211~$\mu$m while the gap between adjacent teeth remains at 9~$\mu$m.
    Compared to the original configuration shown in figure \ref{fig: traps}(d), this design increases the total electrode surface area. Despite this, the total axial heating rate is reduced, illustrating that geometric reconfiguration, rather than surface area minimization alone, can effectively suppress noise. 
    As in previous figures, blue structures indicate RF electrodes and yellow structures indicate DC electrodes.
    }
    \label{fig. heating rate patch in optimized 3D-printed trap}
\end{figure}

Figures \ref{fig: heating mechanism of the 3D-printed trap}(c) and \ref{fig. heating rate patch in optimized 3D-printed trap}(a) compare the spatial distribution of axial heating contributions before and after optimization. 
We observe an overall reduction in the integrated axial heating rate by approximately 1\%. 
While the improvement may appear modest, it validates the principle that strategic geometric control of the electrode layout can positively influence heating performance. 
%This result is particularly encouraging considering the gap width in the present design remains only 9 $\mu$m—significantly smaller than the $\sim$150 $\mu$m full width of the axial heating peak, as shown in figure \ref{fig. heating rate patch in 3D-printed trap}(a). 
This result is particularly encouraging given that the gap width in the present design remains only 9~$\mu$m,
which is significantly smaller than the approximately 150~$\mu$m full width of the axial heating peak, as shown in figure \ref{fig. heating rate patch in 3D-printed trap}(a).
A larger inter-electrode spacing, more commensurate with the width of the heating feature, is expected to yield more substantial reductions.

To support this analysis, figures \ref{fig: traps}(c) and \ref{fig. heating rate patch in optimized 3D-printed trap}(b) illustrate the original and optimized electrode geometries, 
with dimensions clearly annotated. 
The results suggest that future trap designs may benefit from a more systematic incorporation of mode-resolved heating maps into their electrode patterning strategies,
potentially enabling significant improvements in trap performance through purely passive geometric control.

\subsection{Scaling Behavior with Ion-Electrode Distance}

We further investigate the relationship between the ion-to-electrode distance and the axial heating peak location. 
Our simulations, summarized in figure \ref{fig. heating rate patch in 3D-printed trap}(b), show a clear linear dependence: 
as the ion is moved closer to the electrodes, 
the peak heating contribution shifts inward along the axial direction. 
This trend provides a valuable design rule for future electrode engineering efforts and supports the feasibility of tuning trap geometry to suppress specific heating mechanisms.

\section{Discussion}\label{sec6}

The comparative analysis between the conventional blade trap and the novel 3D-printed skeleton trap highlights important insights into the role of electrode geometry in mitigating electric-field noise and reducing ion heating rates in RF Paul traps.

\subsection{ Geometric Influence on Electric Field Noise}

Our results reinforce the hypothesis that electrode surface proximity and geometry critically influence ion heating, especially via fluctuating patch potentials. 
The skeleton architecture of the 3D-printed trap effectively removes large sections of nearby conductive surfaces, particularly those contributing most heavily to the heating rate. 
By spatially displacing noisy regions away from the ion, the skeleton design achieves a heating rate reduction of over 50\% without altering confinement strength.

This structural reconfiguration provides a practical and scalable method to decouple miniaturization from heating performance degradation, 
a long-standing challenge in surface and microfabricated traps \cite{Allcock_2011,Seidelin2006}. 
It opens a promising path toward high-performance, compact quantum devices.

\subsection{Mode-Specific Heating Behavior}

The spatial and directional analysis of field noise further reveals a strong asymmetry between radial and axial mode heating. 
While radial heating is predominantly sourced from the nearest electrode regions, 
axial mode heating exhibits a peak contribution from patches located approximately 110~$\mu$m away. This observation is consistent across both trap designs and is attributed to the anisotropic field vector orientations generated by patch potentials.

This nuanced behavior implies that noise mitigation strategies must account for mode-dependent spatial profiles, 
not just global surface area. 
For example, electrode features that reduce field strength or directional alignment in the axial direction at intermediate distances can be more effective than merely minimizing local surface area.

\subsection{Potential and Limitations of Geometric Optimization}

Our proof-of-concept optimization, which modifies the spacing of the segmented trap electrodes in the 3D-printed trap,
shows that geometric tuning can directly suppress axial heating contributions. 
Though the net reduction achieved was modest (approximately 1\%),
the approach demonstrates that even increased total surface area can result in lower heating, if the high-contribution zones are strategically excluded.

This counterintuitive result suggests that electrode design must focus on noise geometry, 
not just surface minimization. 
However, our optimization was constrained by fabrication limits, 
particularly the narrow 9~$\mu$m tooth gaps compared to the roughly 150~$\mu$m width of the axial heating peak. Future designs with wider electrode spacing or curved contours might yield significantly better performance.

\subsection{Broader Implications for Quantum Information Processing}
The ability to reduce motional heating directly enhances the coherence time and gate fidelity of trapped-ion qubits, 
two critical metrics for fault-tolerant quantum computation \cite{Wineland1998,Schindler_2013}. 
The skeleton trap geometry, 
by enabling low heating with minimal compromise to trap depth and secular frequency,
may serve as a bridge between traditional 3D traps and planar microfabricated arrays, 
combining the best of both architectures.

Furthermore, the linear relationship between ion-electrode distance and axial heating peak location provides a predictive tool for future trap scaling and design optimization, 
potentially facilitating larger and more stable ion trap arrays for scalable quantum processors.

\section{Conclusion}\label{sec7}

In this work, we have presented a theoretical investigation of ion heating in a novel 3D-printed ion trap architecture featuring a skeleton-like electrode structure. By comparing its performance against a conventional blade trap of identical ion-to-electrode distance, we have demonstrated that the skeleton trap achieves a substantial reduction in the total electric-field-induced heating rate, exceeding 50\%, through geometric suppression of field noise from nearby electrode surfaces.

Our spatially resolved analysis revealed that the dominant contributions to heating originate from electrode regions within a few hundred microns of the ion, with axial heating exhibiting a nontrivial peak at an intermediate distance of approximately 110~$\mu$m. This behavior was shown to result from the directional character of electric fields generated by patch potentials, emphasizing the importance of mode-dependent field geometry in trap design.

Furthermore, we explored geometric optimization strategies by adjusting the placement and width of electrode features in the 3D-printed trap. While the initial improvement in total heating rate was modest (around 1\%), the study establishes a foundation for further noise mitigation through design tailoring. Notably, our findings illustrate that reducing heating does not necessarily require reducing total electrode surface area; rather, it requires targeted suppression of high-contribution regions.

By fixing the RF voltage amplitude at a practical value of 150~V and exploring the trap stability space through drive frequency sweeps, we established operating parameters that yield strong confinement while respecting practical and fabrication constraints. These results affirm the feasibility of the proposed architecture for future experimental realization using advanced metal 3D printing technologies.

Overall, this work provides both a physical understanding and a design strategy for reducing motional heating in ion traps.
This represents one of the major obstacles to scaling up trapped-ion quantum computing systems. 
The insights gained here open new avenues for integrating scalable, low-noise ion trap geometries into next-generation quantum information platforms.

\backmatter

%\bmhead{Supplementary information}

%If your article has accompanying supplementary file/s please state so here. 

%Authors reporting data from electrophoretic gels and blots should supply the full unprocessed scans for key as part of their Supplementary information. This may be requested by the editorial team/s if it is missing.

%Please refer to Journal-level guidance for any specific requirements.

\bmhead{Acknowledgements}

We acknowledge support from the National Science and Technology Council (NSTC), Taiwan, under Grant No. NSTC-113-2112-M-002-025 and No. NSTC-112-2112-M-002-001. 
M.-S. Chang acknowledges support from NSTC under Grants No. 112-2112-M-001-070 and 111-2112-M-001-073. 
We are also grateful for valuable discussions with Michael Galvez and Everett Fall of Additive Intelligence regarding various aspects of 3D-printed trap manufacturing.


\bibliography{sn-bibliography}% common bib file

\end{document}